\documentclass[12pt,twoside]{amsart}
\usepackage{mathrsfs,amsthm,amscd,amsmath,amssymb}
\usepackage[colorlinks,linkcolor=green,citecolor=blue,pdfstartview=FitH]{hyperref}
\usepackage{actuarialsymbol}
\usepackage{amsfonts}
\usepackage{graphicx}
\usepackage{caption,subcaption}

\title[Comonotonic and moment matching approximations for sums of lognormal random variables II]
{Comonotonic and moment matching approximations for sums of lognormal random variables II}
\author{Chunle Huang}
\address{Chunle Huang, School of Mathematics, Hunan University, Changsha, 410082, China.}
\email{2019044@hnu.edu.cn; 402961544@qq.com}

\newtheorem{thm}{Theorem}[section]
\newtheorem{lem}[thm]{Lemma}
\newtheorem{cor}[thm]{Corollary}

\theoremstyle{definition}
\newtheorem{defn}[thm]{Definition}
\newtheorem{rem}[thm]{Remark}

\makeatletter
\let\uppercasenonmath\@gobble

\makeatother

\usepackage{booktabs}

\begin{document}
\bibliographystyle{amsalpha+}
\begin{abstract}
In this paper, based on the concept of step-weighted distribution, we introduce a minimum-deformation approach to constructing comonotonic and moment-matching approximations for sums of lognormal random variables. The proposed approach aims to minimize the Pearson deformation of a starting comonotonic approximation subject to moment-matching constraints. We establish several results concerning the existence and uniqueness of minimum-deformation comonotonic approximations and propose a method for determining the step quantiles and weights. In particular, when the number of steps is one greater than the moment-matching order, we show that the joint optimization problem can be reduced to an optimization problem involving only the partition points. The proposed framework accommodates moment matching of any prescribed finite order, and numerical experiments demonstrate its application to approximations matching the first five moments of the original sum. The numerical results also show that minimizing the Pearson deformation does not necessarily improve the approximation accuracy of risk measures, which depends on both the starting comonotonic approximation and the choice of partition points and correction weights.
\end{abstract}
\maketitle
\tableofcontents
\section{Introduction}\label{sec:intro}
In this paper we consider approximations for the distribution and risk measures of a random variable
\[
S=\sum_{i=1}^{n}\alpha_i e^{Z_i}
\]
where \(\alpha_i>0\) are positive real numbers and \((Z_1,\ldots,Z_n)'\) is a multivariate random vector. In practice, the quantities \(\alpha_i\) can be used to denote future deterministic payments or saving amounts, while the random variables \(Z_i\) describe financial risks. Given that Gaussian models are widely used in finance for modeling asset returns, we will assume that \((Z_1,\ldots,Z_n)'\) follows a multivariate normal distribution. In this case, \(S\) becomes a sum of dependent lognormal random variables, and it is generally difficult to derive analytical expressions for its distribution function and risk measures. Therefore, it is of interest to consider its approximations.

In the literature, several techniques for approximating the distribution and risk measures of sums of lognormal random variables have been proposed. One technique is based on the idea of matching the first few moments of \(S\) using tractable distributions. Moment matching approximations are relatively easy to construct and have been widely studied in actuarial science and finance. However, matching a prescribed number of moments does not necessarily guarantee a satisfactory approximation of the distribution of \(S\), particularly in its right tail. For a detailed discussion of the moment matching methods we refer to \cite{KGDD2008, KPW2012, MP1998, MR2000, VHD2005b}.

Another technique is based on the concept of comonotonicity. For a random sum \(S\), the comonotonic sum \(S^c\) provides an upper bound in convex order. On the other hand, for a conditioning random variable \(\Lambda\), the conditional expectation \(S^l=\mathbb E[S\mid\Lambda]\) provides a lower bound in convex order. If the conditional expectations of the individual risks are non-decreasing functions of the same conditioning variable, then \(S^l\) is also a comonotonic sum. In the case of sums of lognormal random variables, the Taylor-based (TB) and maximal-variance (MV) methods provide two important choices of such conditional expectation approximations. The advantage of these comonotonic approximations is that their distribution functions, quantiles and risk measures are relatively easy to calculate. In particular, the additivity of distortion risk measures for comonotonic risks makes this approach attractive in actuarial science and finance. For an extensive overview on the concept of comonotonicity we refer to \cite{DDGK2006, Dhaene et al. (2002a)}. For applications of this concept in actuarial science and finance we refer to \cite{DD2007, Dhaene et al. (2006), DHL2008, DWY2000, VDG2003}. 

Moment matching approximations and comonotonic approximations have their respective advantages. The former reproduce certain moments of the original sum, while the latter preserve a useful dependence structure and facilitate the calculation of risk measures. Following the fundamental work of \cite{Dhaene et al. (2002a), Dhaene et al. (2002b), DVG2005, KDG2000, VCD2008} and based on the concept of step-weighted distribution, we introduced in our previous paper \cite{HuangStep} a class of approximations that are both comonotonic and moment matching. Specifically, starting from a continuous comonotonic approximation, we divided its distribution into several quantile intervals and assigned a nonnegative weight to each interval. By choosing appropriate weights, the resulting step-weighted approximation can match a prescribed number of moments of the original sum. Moreover, the operation of step-weighting is compatible with the operation of comonotonic summing, so that the resulting approximation retains the comonotonic structure. 

Although the step-weighted construction provides a way to obtain comonotonic and moment matching approximations, the choice of the step quantiles and weights deserves further investigation. In our previous paper, the step quantiles were selected in advance, and the corresponding weights were determined by the moment matching conditions. In general, however, the moment matching conditions do not uniquely determine both the step quantiles and weights. Different choices may lead to different step-weighted distributions, even when the same moments are matched. Therefore, a natural question arises as to how the step quantiles and weights should be selected.

To answer this question, we first fix a starting comonotonic approximation \(T\) of the original sum $S$ with good approximation performance. We then modify the distribution of \(T\) using the technique of step-weighting so that the resulting approximation matches the first few moments of \(S\). The compatibility of step-weighting with comonotonic summing ensures that the resulting approximation retains the comonotonic structure. To avoid excessive changes to the distribution of \(T\), we propose a minimum-deformation approach. More precisely, among all step-weighted corrections satisfying the prescribed moment matching conditions, we seek the one that changes the distribution of \(T\) the least. This leads to the problem of determining the step quantiles and weights by minimizing the Pearson deformation subject to the moment matching constraints. Divergence-based optimization under moment constraints has also been investigated in the statistical literature; see, for example, \cite{BKA1998, BMKA2012}.

We first investigate the minimum-deformation problem for a fixed quantile partition. In this case, the moment matching conditions can be expressed as a system of linear equations in the step weights, while the Pearson deformation is a quadratic function of these weights. This allows us to establish theoretical results concerning the existence and uniqueness of the minimum-deformation weights and to obtain explicit representations under suitable conditions. We then allow the quantile partition to vary and consider the joint optimization of the step quantiles and weights. We show that the joint minimum-deformation problem admits an optimal solution whenever its feasible set is nonempty and can be reduced to an optimization problem involving only the step quantiles. Furthermore, we investigate how the minimum Pearson deformation is affected by the number of steps and the number of matched moments. These results provide a theoretical basis for selecting the step quantiles and weights in the construction of comonotonic and moment matching approximations.

It should be pointed out that the minimum-deformation principle does not necessarily lead to better approximation performance. Indeed, minimizing the Pearson deformation only ensures that, among the admissible step-weighted corrections, the selected correction changes the starting distribution as little as possible. It does not guarantee that the resulting approximation is closer to the distribution of \(S\), nor does it necessarily minimize the approximation errors of quantiles or risk measures. In particular, the approximation performance of a minimum-deformation comonotonic approximation may depend substantially on the choice of the starting comonotonic approximation. Thus, minimum deformation and approximation accuracy are two different considerations, and the latter needs to be examined separately. 
However, to obtain a minimum-deformation comonotonic approximation with good approximation performance, it is important to choose a starting comonotonic approximation that already provides a good approximation to the original sum. 

To illustrate the proposed approach, we consider sums of dependent lognormal random variables and use three starting comonotonic approximations: the comonotonic upper bound \(S^c\), the TB lower-bound approximation, and the MV lower-bound approximation. For each starting approximation, we determine the step quantiles and weights by minimizing the Pearson deformation subject to the moment matching constraints. We compare the resulting approximations with their starting counterparts and with the step-weighted approximations based on equally spaced partitions, with particular attention to quantiles and conditional tail expectations. Numerical results show that the minimum-deformation approach can achieve moment matching while preserving the high approximation accuracy of the TB and MV methods. However, the results for \(S^c\) indicate that minimizing the Pearson deformation does not necessarily improve approximation accuracy, particularly in the upper tail. These findings highlight the importance of choosing a good starting comonotonic approximation. Furthermore, using the MV lower-bound approximation as the starting point, we have numerically constructed comonotonic approximations that exactly match the first five moments of the original sum.

The remainder of this paper is organized as follows. Section \ref{103173} collects some preliminary results concerning comonotonicity and step-weighted distributions. Section \ref{sec:min-deformation} is devoted to the study of minimum-deformation theory, including fixed-partition and joint optimization problems. Section \ref{sec:numerical} applies the proposed method to sums of lognormal random variables. Section \ref{1031742} presents numerical illustrations based on different starting comonotonic approximations. Section \ref{1031960} concludes the paper.

\section{Preliminaries} \label{103173}
\subsection{Comonotonicity and convex bounds} Comonotonicity describes a dependence structure in which random variables move together in the same direction. It has been extensively discussed in \cite{Dhaene et al. (2002a), Dhaene et al. (2002b)}. For applications of this concept in finance and actuarial science, the reader is referred to \cite{DDV2011, DVG2005, DWY2000, WD1998}. 
\begin{defn}
A random vector $\underline{X} = (X_1, ..., X_n)$ is said to be comonotonic if $\underline{X} \stackrel{d}{=} (F^{-1}_{X_1}(U), ..., F^{-1}_{X_n}(U))$ where $ \stackrel{d}{=}$ stands for equality in distribution and $U \sim U(0, 1)$.
\end{defn}
The following result is useful when considering comonotonic random vectors. 
\begin{lem} [Theorem 3 in \cite{Dhaene et al. (2002a)}] \label{626115} 
A random vector $\underline{X}$ is comonotonic if and only if there exist a random variable $Z$ and non-decreasing functions $f_i, i = 1, ..., n$, such that $\underline{X} \stackrel{d}{=} (f_1(Z), ..., f_n(Z)).$ \end{lem}
Let $S = \sum^n_{i = 1}X_i$ and $S^c = \sum^n_{i = 1}X_i^c$, where $(X_1^c, ..., X_n^c)$ is a comonotonic modification of $(X_1, ..., X_n)$. 
For any conditioning random variable $\Lambda$ let $S^l = \sum^n_{i = 1}E[X_i \,|\, \Lambda].$ Note that if there exists a conditioning random variable $\Lambda$ such that all the conditional expectations $E[X_i \,|\, \Lambda], i = 1, ..., n$, are non-decreasing functions of $\Lambda$ then $S^l$ will be a comonotonic sum. We introduce the following result, concerning convex order bounds for sums of dependent random variables. 
\begin{thm} [Propositions 2 and 5 in \cite{KDG2000}] \label{65401} For any integrable random variables $X_1,\ldots,X_n$ and any conditioning random variable $\Lambda$, we have $S^l\leq_{cx}S\leq_{cx}S^c.$ 
\end{thm}
From Theorem \ref{65401} $S^c$ and $S^l$ give convex bounds for the original sum $S$. On the other hand, in case the dependence between the random variables $X_1, ..., X_n$ is unknown or too cumbersome to work with so that it is impossible to derive the distribution function of $S$ exactly, we can regard $S^c$ and $S^l$ as approximations to $S$. For a detailed discussion of the random sums $S^c$ and $S^l$ as approximations to $S$ and their applications in actuarial science and finance we refer to \cite{Dhaene et al. (2002a), Dhaene et al. (2002b), KDG2000, VCD2008}. 
\subsection{Step-weighted distribution and moment matching approximations}\label{subsec:step-theory} 
The concept of weighted distribution can be traced back to the late 1970s in the statistical literature, see the original work of Patil, Ord and Rao in \cite{PR1976} and \cite{PR1978}, and has proven to be a useful tool in the study of risk measures after the pioneering work by Furman and Landsman in \cite{FL2005} and \cite{FL2008} and Furman and Zitikis in \cite{FZ2008a}, \cite{FZ2008b} and \cite{FZ2009}. The reader is also referred to the recent work of Denuit in \cite{Denuit2019} and \cite{Denuit2020} for a related introduction to the weighted transform and its properties. In this paper, we will consider the case where the weight function is a step function and the corresponding weight distribution will be called step-weighted distribution as discussed in \cite{HuangStep, HuangDRM}.

\begin{defn}\label{52378}
Let \(A=(a_1,\ldots,a_m)'\) and \(Q=(q_1,\ldots,q_m)'\) be two numerical vectors with 
\(\min(a_1,\ldots,a_m)\geq 0, \max(a_1,\ldots,a_m)>0\) and
\(
0=q_0 < q_1 < \cdots < q_m = 1.
\)
Let $X$ be a continuous random variable, that is, its distribution function $F_X$ is continuous on $(-\infty,\infty)$. Then we call $\omega_{A,Q,X}$ the step-weight function of $X$ associated with $A$ and $Q$, abbreviated as $\omega_{A,Q}$ when there is no ambiguity, if 
$$\omega_{A,Q,X}(x)=a_k, \,\, F_X^{-1}(q_{k-1})\le x<F_X^{-1}(q_k), k=1,\ldots,m$$ 
where, for $k=1$, the left endpoint $F_X^{-1}(q_0)$ is replaced by $F_X^{-1+}(q_0)$ and for $k=m$, the right endpoint is included, and $\omega_{A,Q,X}(x)=0$ outside these intervals. The step-weight function $\omega_{A,Q,X}$ is called \emph{normalized} if \(\sum_{j=1}^{m} a_j(q_j-q_{j-1})=1\). We call $X_{A,Q}$ the step-weighted version of $X$ associated with $A$ and $Q$ if 
\[dF_{X_{A,Q}}(x) = \frac{\omega_{A, Q, X}(x)}{E[\omega_{A, Q, X}(X)]}\,dF_X(x), \,\, x\in\mathbb{R}.\]
\end{defn}
The step-weighted distribution redistributes the probability mass of \(X\) among its quantile intervals. In Definition \ref{52378}, adjacent quantile intervals are allowed to carry the same weight, and whether such intervals are kept separate or merged has no effect on the resulting step-weighted distribution. Since multiplying all step heights by the same positive constant does not change the step-weighted distribution, \emph{we henceforth restrict attention to normalized step-weight functions}, so that $\sum_{j=1}^{m} a_j\Delta q_j=1$ where $\Delta q_j=q_j-q_{j-1}$. The following result establishes an important relation between step-weighted distributions and comonotonic sums.
\begin{thm} [Theorem 5.1 in \cite{HuangStep}] \label{1024148}
Let $A$ and $Q$ be given as in Definition \ref{52378}. Then for any continuous random variables $X_i$ with step-weighted versions $X_{i, A, Q}$ 
$$(X_1^c + \cdots + X_n^c)_{A, Q} \overset{d}{=} (X_{1, A,Q})^c+\cdots+(X_{n, A,Q})^c.$$
\end{thm}

Consider the sum $S = \sum^n_{i = 1}X_i$ of continuous random variables $X_1, ..., X_n$ and let $T$ be a continuous comonotonic approximation of $S$, which means that there exist continuous random variables $Y_1, ..., Y_n$ such that $T = Y^c_1 + \cdots + Y^c_n.$ Assume that \(S\) and \(T\) have finite raw moments up to order \(r\). Define $$B_{r+1,m}(Q) = (M_{kj}(Q))_{k=0,\ldots,r;\,j=1,\ldots,m} \,\, \text{and} \,\, c_{r+1} = (1,E[S],\ldots,E[S^r])' $$ where $$M_{kj}(Q) = \int_{q_{j-1}}^{q_j} [F_T^{-1}(u)]^k\,du, \,\, k=0,1,\ldots,r,  j=1,\ldots,m.$$ Then, one can verify that the normalization condition together with the first $r$ moment-matching conditions given by Theorem 6.1 in \cite{HuangStep} can be written as 
\begin{equation} \label{1031122}
B_{r+1,m}(Q)A=c_{r+1} \,\, \text{and} \,\, A\geq 0.
\end{equation}
Any solution of system (\ref{1031122}) defines a comonotonic and moment-matching approximation of $S$; see Theorem 6.1 in \cite{HuangStep}. However, for a fixed quantile partition, system (\ref{1031122}) may admit multiple nonnegative solutions, while different partitions may also lead to different moment-matching approximations. Thus, in general, moment matching alone does not uniquely determine a comonotonic and moment-matching approximation. This motivates the selection principle developed in the next section.

\section{Minimum-deformation theory}
\label{sec:min-deformation}
As shown in the preceding section, moment matching alone does not, in general, uniquely determine a step-weighted comonotonic approximation. Given a starting comonotonic approximation \(T\) of \(S\), we seek, among all step-weighted corrections satisfying the prescribed moment matching conditions, the one that changes the distribution of \(T\) as little as possible. This leads to the minimum-deformation principle developed in this section. Throughout this section, let $S$ and $T$ be continuous random variables with finite raw moments up to order $r$, where $r$ is a positive integer. Also, let $m$ be a positive integer denoting the number of quantile intervals when considering step-weighted distributions.

\subsection{Minimum-deformation principle}
To implement the minimum-deformation principle, we need to quantify the departure of a step-weighted correction from its starting distribution. A natural measure for this purpose is the Pearson divergence 
\[
D(F_{T_{A,Q}},F_T)=\int\left(\frac{dF_{T_{A,Q}}}{dF_T}-1\right)^2\,dF_T 
\]
which belongs to the general class of divergences studied by Ali and Silvey \cite{AS1966}; see also Liese and Vajda \cite{LV2006}. The following lemma shows that for step-weighted distributions, this divergence takes a particularly simple form, reducing the distributional deformation to a quadratic function of the step heights.

\begin{lem} \label{lem:pearson}
Let $T$ be a continuous random variable and $T_{A,Q}$ be a normalized step-weighted correction of $T$ associated with $A$ and $Q$. Then it holds that 
\[D(F_{T_{A,Q}},F_T)=\sum_{j=1}^m \Delta q_j(a_j-1)^2.\]
Hence $D_P\ge0$, with equality if and only if the corrected and starting laws coincide.
\end{lem}
\begin{proof}
On the $j$th quantile interval the Radon--Nikodym derivative equals $a_j$ and the interval has $F_T$-probability $\Delta q_j$. Direct integration gives the result.
\end{proof}

\begin{defn} \label{def:MD}
Denote by $\mathcal{F}_{r,m}$ the set of all pairs $(A,Q)$ defining normalized step-weighted versions of $T$ that match the first $r$ raw moments of $S$. A step-weighted version $T_{A^*,Q^*}$, with $(A^*,Q^*)\in\mathcal{F}_{r,m}$, is called a \emph{minimum-deformation approximation} of $S$ relative to $T$ if
\begin{equation} \label{1016198}
D(F_{T_{A^*,Q^*}},F_T)=\inf_{(A,Q)\in\mathcal{F}_{r,m}}D(F_{T_{A,Q}},F_T).
\end{equation}
\end{defn}

Notice that the infimum in Definition \ref{def:MD} need not be attained when the partition is allowed to vary. Indeed, a sequence of partitions may approach a degenerate partition in which one or more quantile intervals have vanishing widths. Thus the limiting correction need not belong to the class $\mathcal{F}_{r, m}$. This explains the use of an infimum in Definition \ref{def:MD}. Later, when studying variable partitions, we impose a positive lower bound on the interval widths in order to recover compactness and establish existence.

Moment matching specifies what the correction must reproduce; minimum deformation determines which correction is preferred among those satisfying the moment conditions. Thus, the two principles play complementary roles: moment matching determines the class of candidate corrections, while minimum deformation provides a criterion for selecting among them. Using the notation of Subsection \ref{subsec:step-theory} and Lemma \ref{lem:pearson}, the optimization problem (\ref{1016198}) can be written as 
\begin{equation} \label{102180}
\inf_{A,Q}\sum_{j=1}^{m}\Delta q_j(a_j-1)^2 \,\, \text{subject to} \,\, B_{r+1,m}(Q)A=c_{r+1} \,\, \text{and} \,\, A\geq 0.
\end{equation}
\subsection{Fixed-partition minimum deformation} \label{1017211}
In this subsection we fix a partition $Q=(q_1,\ldots,q_m)'$ of the unit interval such that $0=q_0 < q_1 < \cdots < q_m = 1$. Define $W(Q)=\operatorname{diag}(\Delta q_1,\ldots,\Delta q_m)$, where $\Delta q_i = q_i - q_{i - 1}, i = 1, ..., m$, and 
\[
 \mathcal A^{\mathrm{feas}}_{r+1,m}(Q)=\{A\in\mathbb R^m:A\ge0,\ B_{r+1,m}(Q)A=c_{r+1}\}.
\]

\begin{thm} \label{thm:fixed}
Assume that the feasible weight set $\mathcal A^{\mathrm{feas}}_{r+1,m}(Q)$ is nonempty. Then there exists a unique vector $A^*(Q)\in\mathcal A^{\mathrm{feas}}_{r+1,m}(Q)$ minimizing the Pearson deformation \(D_Q(A)=(A-{\bf1})'W(Q)(A-{\bf1}).\) Furthermore, if $B_{r+1,m}(Q)$ has full row rank, the unique minimizer over the affine space $\left\{ A \in \mathbb{R}^m : B_{r+1,m}(Q)A = c_{r+1} \right\}$ is given by 
\begin{equation}\label{eq:explicit}
 A(Q)= {\bf1}+P_{r + 1, m}(Q)[c_{r+1}-B_{r+1,m}(Q){\bf1}].
\end{equation}
where $P_{r + 1, m}(Q)=W(Q)^{-1}B_{r+1,m}(Q)'\left[B_{r+1,m}(Q)W(Q)^{-1}B_{r+1,m}(Q)'\right]^{-1}$. In this case, $A^*(Q)=A(Q)$ whenever $A(Q)\ge0$.
\end{thm}

\begin{proof}
For fixed $Q$ with $0=q_0<q_1<\cdots<q_m=1$, every $\Delta q_j$ is positive, so $W(Q)$ is positive definite. Hence $D_Q$ is strictly convex. The set $\mathcal A^{\mathrm{feas}}_{r+1,m}(Q)$ is closed and convex. It is also bounded: the normalization equation is the first row of $B_{r+1,m}(Q)A=c_{r+1}$ and gives $\sum_{j=1}^{m}\Delta q_j a_j=1.$  Since $a_j\ge0$, we have
\[
 0\le a_j\le{1}/{\Delta q_j}, \quad j=1,\ldots,m.
\]
Therefore, $\mathcal A^{\mathrm{feas}}_{r+1,m}(Q)$ is compact. Since $\mathcal A^{\mathrm{feas}}_{r+1,m}(Q)$ is nonempty and $D_Q$ is continuous, a minimizer exists. Since $\mathcal A^{\mathrm{feas}}_{r+1,m}(Q)$ is convex and $D_Q$ is strictly convex, the minimizer is unique. For the second part, we temporarily ignore the nonnegativity constraint $A \ge 0$ and minimize over the affine space $\left\{ A \in \mathbb{R}^m : B_{r+1,m}(Q)A = c_{r+1} \right\}$. To this end, we introduce a Lagrange multiplier $\lambda \in \mathbb{R}^{r+1}$ and write 
\[
 L(A,\lambda)
 =
 (A-{\bf1})'W(Q)(A-{\bf1})
 +2\lambda'[B_{r+1,m}(Q)A-c_{r+1}].
\]
The first-order condition with respect to $A$ is $W(Q)(A-{\bf1})+B_{r+1,m}(Q)'\lambda=0.$ Thus
\[
 A={\bf1}-W(Q)^{-1}B_{r+1,m}(Q)'\lambda.
\]
Substituting this expression into $B_{r+1,m}(Q)A=c_{r+1}$ gives
\[
 B_{r+1,m}(Q)W(Q)^{-1}B_{r+1,m}(Q)'\lambda
 =
 B_{r+1,m}(Q){\bf1}-c_{r+1}.
\]
Full row rank of $B_{r+1,m}(Q)$ implies that the matrix on the left is positive definite and therefore invertible. Substituting the solution for $\lambda$ into the expression for $A$ yields \eqref{eq:explicit}. If the resulting vector is nonnegative, it is feasible for the original problem and it must be the unique constrained minimizer.
\end{proof}

Theorem \ref{thm:fixed} gives a useful two-level description of the fixed-partition problem: existence and uniqueness. Existence follows from compactness of the feasible set and uniqueness from strict convexity of the Pearson deformation. Theorem \ref{thm:fixed} also separates two computational regimes. If the affine optimizer \eqref{eq:explicit} is nonnegative, the correction is explicit. If one or more components are negative, the affine optimizer is no longer feasible and the nonnegativity constraints affect the solution. In this case, the minimizer still exists and is unique, but numerical methods may be required to determine it. Thus, failure of the closed-form solution is not failure of the minimum-deformation principle. 

\subsection{Joint minimum deformation} \label{919664}
Although the fixed-partition problem has a unique solution whenever its feasible set is nonempty, different partitions may yield different minimum deformations. We therefore consider in this subsection the joint minimum-deformation problem, allowing both the partition and the step weights to vary. To ensure compactness of the partition set and exclude degenerate intervals, fix $\delta>0$ such that $m\delta\leq 1$ and denote 
$$\mathcal{Q}_{m,\delta}=\left\{Q : 0=q_0<q_1<\cdots<q_m=1, \Delta q_j\geq\delta, j = 1, ..., m\right\}$$ 
where $Q=(q_1,\ldots,q_m)'$. 
Define the joint feasible set $\mathcal{J}_{r+1,m,\delta}^{\mathrm{feas}}$ by
\[
\mathcal{J}_{r+1,m,\delta}^{\mathrm{feas}}=\left\{(Q,A):Q\in\mathcal{Q}_{m,\delta}, A\geq 0, B_{r+1,m}(Q)A=c_{r+1}\right\}
\]
and the feasible partition set $\mathcal{Q}_{m,\delta}^{\mathrm{feas}}$ by
\[
\mathcal{Q}_{m,\delta}^{\mathrm{feas}}=\left\{Q\in\mathcal{Q}_{m,\delta}:\text{there exists }A\geq 0\text{ such that }B_{r+1,m}(Q)A=c_{r+1}\right\}.
\]
The joint minimum-deformation problem is then formulated as
\begin{equation} \label{1016266}
\min_{(Q,A)\in\mathcal{J}_{r+1,m,\delta}^{\mathrm{feas}}}\sum_{j=1}^{m}\Delta q_j(a_j-1)^2.
\end{equation}
\begin{thm}\label{thm:existence}
Assume that the joint feasible set $\mathcal{J}_{r+1,m,\delta}^{\mathrm{feas}}$ is nonempty. Then the joint minimum-deformation problem (\ref{1016266}) admits at least one optimal solution. For each $Q\in\mathcal{Q}_{m,\delta}^{\mathrm{feas}}$, let $A^*(Q)$ be the unique fixed-partition minimizer established in Theorem \ref{thm:fixed}, and define $\Psi(Q)=D_Q(A^*(Q))$. Then, problem (\ref{1016266}) has a unique optimal solution if and only if $\Psi$ has a unique minimizer $Q^*$ over $\mathcal{Q}_{m,\delta}^{\mathrm{feas}}$. In this case, the optimal solution is $(Q^*,A^*(Q^*))$.
\end{thm}

\begin{proof}
For \(k=1,\ldots,r\), define 
\(
H_k(q)
=
\int_0^q (F_T^{-1}(u))^k\,du.
\)
The finite-moment assumption implies that
$(F_T^{-1})^k$ is integrable on $(0,1)$, and hence
$H_k$ is absolutely continuous, in particular continuous, on $[0,1]$. Hence
\(
M_{kj}(Q)
=
H_k(q_j)-H_k(q_{j-1})
\)
is continuous in $Q$, and therefore $B_{r+1,m}(Q)$ is continuous in $Q$.
The set $\mathcal{Q}_{m,\delta}$ is compact. For every feasible pair,
normalization gives $\sum_{j=1}^m \Delta q_j a_j=1.$ Since $\Delta q_j\geq\delta$ and $a_j\geq0$, we have
\[
0\leq a_j\leq 1/\delta,\quad j=1,\ldots,m.
\]
Moreover, since $B_{r+1,m}(Q)$ is continuous in $Q$, the mapping
\(
(Q,A) \mapsto B_{r+1,m}(Q)A
\) 
is continuous. Hence the constraint set
\(
\{(Q,A):B_{r+1,m}(Q)A=c_{r+1}\}
\)
is closed. Thus $\mathcal{J}_{r+1,m,\delta}^{\mathrm{feas}}$ is a closed subset of the compact set
\(
\mathcal{Q}_{m,\delta}\times[0,1/\delta]^m,
\)
and is therefore compact. Note that the objective in (\ref{1016266}) is continuous. The Weierstrass theorem therefore gives at least one minimizer. For each $Q\in\mathcal{Q}_{m,\delta}^{\mathrm{feas}}$, Theorem \ref{thm:fixed} guarantees a unique fixed-partition minimizer $A^*(Q)$. Consequently, the joint problem is equivalent to minimizing $\Psi(Q)$ over $\mathcal{Q}_{m,\delta}^{\mathrm{feas}}$. If $\Psi$ has a unique minimizer $Q^*$, then $(Q^*,A^*(Q^*))$ is the unique optimal solution to problem (\ref{1016266}). Conversely, if $\Psi$ has two distinct minimizers $Q_1$ and $Q_2$, then $(Q_1,A^*(Q_1))$ and $(Q_2,A^*(Q_2))$ are two distinct optimal solutions to problem (\ref{1016266}). Thus the joint optimal solution is unique if and only if $\Psi$ has a unique minimizer.
\end{proof}

Theorem \ref{thm:existence} provides a theoretical basis for quantile selection in the minimum-deformation framework. The partition $Q$ determines where the weighting is applied, whereas the vector $A$ determines the corresponding weights. Thus, rather than prescribing the quantile partition in advance and optimizing only the weights, the joint minimum-deformation problem determines $Q$ and $A$ simultaneously subject to the exact moment-matching constraints. The restriction $\Delta q_j\geq\delta$ excludes degenerate intervals and ensures compactness of the admissible partition class. Thus, the joint optimization determines both where the weighting is applied and the corresponding weights, thereby providing an optimization-based criterion for quantile selection.

\begin{rem}
The reduced objective function $\Psi$ admits an explicit representation whenever $B_{r+1,m}(Q)$ has full row rank and the affine optimizer is nonnegative. Under these conditions, Theorem \ref{thm:fixed} gives
\[
A^*(Q)
=\mathbf{1}+P_{r + 1, m}(Q)\bigl(c_{r+1}-B_{r+1,m}(Q)\mathbf{1}\bigr) 
\]
and hence
\[
\Psi(Q)
=
\bigl(c_{r+1}-B_{r+1,m}(Q)\mathbf{1}\bigr)'
\bigl[B_{r+1,m}(Q)W(Q)^{-1}B_{r+1,m}(Q)'\bigr]^{-1}
\bigl(c_{r+1}-B_{r+1,m}(Q)\mathbf{1}\bigr).
\]
However, the minimization of $\Psi(Q)$ over the feasible partition set $\mathcal{Q}_{m,\delta}^{\mathrm{feas}}$ generally does not admit a closed-form solution. The reason is that $B_{r+1,m}(Q)$ depends nonlinearly on the partition points through the quantile integrals, while $W(Q)$ varies with $Q$ through the step lengths. Moreover, the matrix inverse appearing in the reduced objective further complicates the dependence on $Q$. Therefore, although the optimal correction weights \(A^*(Q)\) admit an explicit representation under the stated conditions, the optimal partition \(Q^*\) generally needs to be determined by numerical optimization.
\end{rem}

\subsection{Selection of the number of steps}\label{subsec:step-selection}
In this subsection, we investigate the selection of the number of steps $m$ by examining the rank structure of the moment matrix and the solvability of the associated moment-matching system. The following result shows that, under the continuity assumption imposed on the starting approximation $T$, this moment matrix always has the largest rank permitted by its dimensions.

\begin{thm}\label{thm:moment-matrix-rank}
Let $T$ be a continuous random variable with finite raw moments up to order $r$, and let $Q=(q_1,\ldots,q_m)'$ be a partition of the unit interval satisfying $0=q_0<q_1<\cdots<q_m=1.$ Then 
\begin{equation} \label{1017336}
\operatorname{rank} B_{r+1,m}(Q)=\min\{r+1,m\}.
\end{equation}
More precisely, if $p=\min\{r+1,m\}$ and $1\leq j_1<\cdots<j_p\leq m$, then the $p\times p$ submatrix formed by the first $p$ rows and the columns $j_1,\ldots,j_p$ has strictly positive determinant. Consequently, the following statements hold: 
\begin{enumerate}
\item $B_{r+1,m}(Q)$ has full row rank if and only if $m\ge r+1$;
\item $B_{r+1,m}(Q)$ has full column rank if and only if $m\le r+1$;
\item $B_{r+1,m}(Q)$ is nonsingular if and only if $m=r+1$.
\end{enumerate}
\end{thm}
\begin{proof}
Let $f(u)=F_T^{-1}(u)$, $I_j=(q_{j-1},q_j), j = 1, ..., m$, and $p=\min\{r+1,m\}.$ For fixed indices $1\le j_1<\cdots<j_p\le m,$ consider the $p\times p$ submatrix
\[
C=
\left(
\int_{I_{j_\ell}} f(u)^k\,du
\right)_{\substack{k=0,\ldots,p-1; \ell=1,\ldots,p}}.
\]
Since $p-1\le r$, all entries of $C$ are finite. By multilinearity of the determinant in its columns,
\[
\det C
=
\int_{I_{j_1}}\cdots\int_{I_{j_p}}
\det
\begin{pmatrix}
1 & \cdots & 1\\
f(u_1) & \cdots & f(u_p)\\
\vdots & & \vdots\\
f(u_1)^{p-1} & \cdots & f(u_p)^{p-1}
\end{pmatrix}
\,du_p\cdots du_1.
\]
The determinant in the integrand is the Vandermonde determinant. Hence, 
\[
\det C
=
\int_{I_{j_1}}\cdots\int_{I_{j_p}}
\prod_{1\le s<t\le p}
\bigl[f(u_t)-f(u_s)\bigr]
\,du_p\cdots du_1.
\]
Because $j_1<\cdots<j_p$, for every $u_\ell\in I_{j_\ell}, \ell=1,\ldots,p$, we have $u_1<\cdots<u_p.$ Since the distribution function of $T$ is continuous, its quantile function $f=F_T^{-1}$ is strictly increasing on $(0,1)$. Therefore $f(u_1)<\cdots<f(u_p),$ and hence
\[
\prod_{1\le s<t\le p}
\bigl[f(u_t)-f(u_s)\bigr]>0
\]
throughout $I_{j_1}\times\cdots\times I_{j_p}$. In particular, the integrand is nonnegative and measurable. By Tonelli's theorem,
\[
\det C
=
\int_{I_{j_1}\times\cdots\times I_{j_p}}
\prod_{1\le s<t\le p}
\bigl[f(u_t)-f(u_s)\bigr]
\,du_1 \ldots du_p.
\]
Since each interval $I_{j_\ell}$ has positive length, the product set $I_{j_1}\times\cdots\times I_{j_p}$ has positive Lebesgue measure. As the integrand is strictly positive throughout this set, it follows that $\det C>0.$ Thus $B_{r+1,m}(Q)$ contains a nonsingular $p\times p$ submatrix, and hence
\[
\operatorname{rank}B_{r+1,m}(Q)\ge p.
\]
On the other hand, since $B_{r+1,m}(Q)$ has $r+1$ rows and $m$ columns,
\[
\operatorname{rank}B_{r+1,m}(Q)
\le
\min\{r+1,m\}.
\]
Therefore, 
\[
\operatorname{rank}B_{r+1,m}(Q) = \min\{r+1,m\}.
\]
The remaining assertions follow immediately from the rank formula (\ref{1017336}). 
\end{proof}

\begin{cor} \label{109763}
Under the assumptions of Theorem \ref{thm:moment-matrix-rank} consider the moment-matching system $B_{r+1,m}(Q)A=c_{r+1}.$ Then the following statements hold:
\begin{enumerate}
\item If $m<r+1$, the system has a solution if and only if $c_{r+1}$ belongs to the column space of $B_{r+1,m}(Q)$. It is unique whenever a solution exists.
\item If $m=r+1$, the system has the unique solution $A=B_{r+1,r+1}(Q)^{-1}c_{r+1}.$ 
\item If $m>r+1$, the system is always solvable, and its solution set is an affine space of dimension $m-r-1$.
\end{enumerate}
In all three cases, a nonnegative moment-matching weight vector exists if and only if the solution set intersects the nonnegative orthant. In particular, when $m=r+1$, this condition reduces to $B_{r+1,r+1}(Q)^{-1}c_{r+1}\ge0.$ 
\end{cor}

\begin{proof}
By Theorem \ref{thm:moment-matrix-rank}, \(\operatorname{rank}B_{r+1,m}(Q)=\min\{r+1,m\}.\) If $m<r+1$, the matrix has full column rank, so the system has at most one solution, which exists precisely when $c_{r+1}$ belongs to its column space. If $m=r+1$, the matrix is nonsingular, giving the stated unique solution. If $m>r+1$, the matrix has full row rank, so the system is always solvable for every $c_{r+1}$. Its solution set is an affine translate of the null space, whose dimension is $m-r-1$. The final assertions follow by imposing the nonnegativity restriction on the weight vector.
\end{proof}

We now address the problem of how the number of steps $m$ should be chosen relative to the number of moments $r$ to be matched. Here, $m$ refers to the number of quantile intervals entering the moment-matching system. Theorem \ref{thm:moment-matrix-rank} and Corollary~\ref{109763} distinguish three regimes. When $m<r+1$, the moment-matching system need not be solvable. When $m>r+1$, the system has $m-r-1$ additional weight degrees of freedom for each fixed partition. In contrast, when $m=r+1$, the moment matrix is square and nonsingular, and the weight vector is uniquely determined by the moment-matching conditions. Thus, $m=r+1$ is the unique dimension-matching specification in which the number of step weights equals the number of independent equality constraints, with no redundant weight degrees of freedom. For this reason, we adopt $m=r+1$ as a canonical specification when a separate tuning of the number of step intervals is not desired. This is a structural design principle, not a claim that $m=r+1$ minimizes the deformation over all possible values of $m$.

An additional advantage of this choice is that it separates the roles of moment matching and minimum deformation. For each fixed partition $Q$, the moment-matching conditions uniquely determine the weight vector $A(Q)=B_{r+1,r+1}(Q)^{-1}c_{r+1}$. Provided that $A(Q)$ is nonnegative, the remaining task is to select, among the feasible quantile partitions, one that minimizes the deformation. Thus, the joint optimization over the partition and the weights reduces to an optimization over the $r$ interior quantile points. This reduction eliminates the weight vector from the optimization variables and provides a direct connection between the minimum-deformation principle and the selection of quantile partitions. Moreover, the rule $m=r+1$ determines the number of quantile intervals directly from the prescribed moment order, without introducing an additional tuning parameter. Allowing $m$ to vary would introduce a separate model-selection problem, potentially requiring an explicit trade-off between deformation and model complexity, which lies beyond the scope of the present study.

We now consider the implications of the choice $m=r+1$ for the two minimum-deformation problems introduced in Subsections \ref{1017211} and \ref{919664}. Recall that, for a fixed partition \(Q=(q_1,\ldots,q_m)'\) of the unit interval with $0=q_0< q_1 < \cdots<q_m=1$, the fixed-partition minimum-deformation problem is 
\begin{equation} \label{1010790}
\min_{A\geq 0} D_Q(A) \,\, \text{subject to} \,\, B_{r+1,m}(Q)A=c_{r+1},
\end{equation}
where $D_Q(A)=(A-\mathbf{1})'W(Q)(A-\mathbf{1})$. 
Allowing the partition $Q$ to vary over the admissible partition set $\mathcal{Q}_{m,\delta}$ 
leads to the joint minimum-deformation problem 
\begin{equation} \label{1010798}
\min_{Q\in\mathcal{Q}_{m,\delta},\,A\geq 0} D_Q(A) \,\, \text{subject to} \,\, B_{r+1,m}(Q)A=c_{r+1}.
\end{equation}
The choice $m=r+1$ leads to significant simplifications of both problems, as stated in the following two theorems. 

\begin{thm} \label{1010803}
Let $m=r+1$ and fix a partition \(Q=(q_1,\ldots,q_{r + 1})'\) of the unit interval satisfying $0=q_0<q_1<\cdots<q_{r+1}=1$. Then, the fixed-partition minimum-deformation problem (\ref{1010790}) admits a solution if and only if \(A(Q) =B_{r+1,r+1}(Q)^{-1}c_{r+1}\geq 0.\) In this case, it has a unique minimizer which is given by $A^*=A(Q)$.  
\end{thm}

\begin{proof}
Recall that for the fixed partition $Q$, the feasible weight set is given by
\[
\mathcal{A}^{\mathrm{feas}}_{r+1,r+1}(Q)
=\left\{A\in\mathbb{R}^{r+1}:A\geq 0, B_{r+1,r+1}(Q)A=c_{r+1}\right\}.
\]
By Theorem \ref{thm:moment-matrix-rank}, the matrix $B_{r+1,r+1}(Q)$ is nonsingular. Hence the moment-matching equations $B_{r+1,r+1}(Q)A=c_{r+1}$ have the unique solution $A(Q)=B_{r+1,r+1}(Q)^{-1}c_{r+1}$. Consequently, $\mathcal{A}^{\mathrm{feas}}_{r+1,r+1}(Q)$ is nonempty if and only if $A(Q)\geq 0$, in which case $\mathcal{A}^{\mathrm{feas}}_{r+1,r+1}(Q)=\{A(Q)\}$. By Theorem \ref{thm:fixed}, the fixed-partition minimum-deformation problem (\ref{1010790}) admits a unique minimizer, necessarily given by $A^*=A(Q)$, with minimum deformation $D_Q(A^*) =(A^*-\mathbf{1})'W(Q)(A^*-\mathbf{1})$. This completes the proof.
\end{proof}

Theorem \ref{1010803} shows that, when \(m=r+1\), the moment-matching conditions uniquely determine the correction weights for each fixed partition. Consequently, the minimum-deformation principle does not play a role in selecting the weights; it becomes relevant only when the partition is allowed to vary. Moreover, one can check that provided that the weights are strictly positive, small perturbations of the prescribed moments lead to correspondingly small changes in the correction weights. Thus, the choice \(m=r+1\) not only eliminates redundant weight degrees of freedom but also provides a locally stable relationship between the prescribed moments and the resulting correction weights.

\begin{thm}\label{1010849}
Let $m=r+1$ and fix $\delta>0$ such that $(r+1)\delta\leq 1$. Then, the joint minimum-deformation problem (\ref{1010798}) admits a solution if and only if $\mathcal Q_{r+1,\delta}^{\mathrm{feas}}\neq\varnothing$. In this case, problem (\ref{1010798}) is equivalent to the reduced problem 
\begin{equation} \label{1010852}
\min_{Q\in\mathcal Q_{r+1,\delta}^{\mathrm{feas}}}(B_{r+1,r+1}(Q)^{-1}c_{r+1}-\mathbf{1})'W(Q)(B_{r+1,r+1}(Q)^{-1}c_{r+1}-\mathbf{1}).
\end{equation}
\end{thm}

\begin{proof}
By Theorem \ref{thm:moment-matrix-rank}, the moment-matching equations uniquely determine \(A(Q)=B_{r+1,r+1}(Q)^{-1}c_{r+1}\) for each fixed partition \(Q\). Hence, a pair \((Q,A)\) satisfies the constraints of the joint minimum-deformation problem (\ref{1010798}) if and only if \(Q\in\mathcal Q_{r+1,\delta}^{\mathrm{feas}}\) and \(A=A(Q)\). Consequently,
\[
\mathcal J_{r+1,r+1,\delta}^{\mathrm{feas}}
=\{(Q,A(Q)):Q\in\mathcal Q_{r+1,\delta}^{\mathrm{feas}}\}.
\]
Thus, the joint feasible set $\mathcal J_{r+1,r+1,\delta}^{\mathrm{feas}}$ is nonempty if and only if \(\mathcal Q_{r+1,\delta}^{\mathrm{feas}}\ne\varnothing\). Whenever this condition holds, the existence of an optimal pair follows from Theorem \ref{thm:existence}. Substituting \(A(Q)\) into the objective function yields the reduced problem (\ref{1010852}). Therefore, any minimizer \(Q^*\) of (\ref{1010852}) determines an optimal pair \((Q^*,A(Q^*))\) of (\ref{1010798}).
\end{proof}

Theorem \ref{1010849} further clarifies the role of the choice \(m=r+1\) in the joint minimum-deformation framework. In this case, the moment-matching conditions uniquely determine the correction weights for each fixed partition, so the joint optimization over partitions and weights reduces to an optimization over the partition alone. Thus, moment matching determines the weights, whereas minimum deformation selects an optimal quantile partition. The nonnegativity requirement is incorporated into the feasible partition set, and the resulting reduced problem admits at least one optimal partition whenever this set is nonempty. Although the correction weights are uniquely determined for each fixed partition, the optimal partition itself need not be unique. Consequently, the uniqueness of the joint minimum-deformation solution depends entirely on the uniqueness of the minimizer of the reduced problem. 

\begin{thm} \label{1016884}
Consider the random variable $S$ and its approximation $T$. Assume that \(S\) and \(T\) have finite second moments and that $F_T$ is continuous on $\mathbb{R}$ and strictly increasing on $(L,R)$ where $L=\operatorname{ess\,inf}T$ and $R=\operatorname{ess\,sup}T$. Define
\[
V_{\max}=
\begin{cases}
(\mathbb{E}[T]-L)(R-\mathbb{E}[T]), & L,R\in\mathbb{R},\\
+\infty, & \text{otherwise}.
\end{cases}
\]
If $\mathbb{E}[S]=\mathbb{E}[T]$ and $0<\operatorname{Var}(S)<V_{\max}$, then there exists $\delta_0>0$ such that
\( 
\mathcal{Q}_{3,\delta}^{\mathrm{feas}}\neq\varnothing
\) 
for every $0<\delta\leq\delta_0$. Consequently, the joint minimum-deformation problem (\ref{1010798}) with $r=2$ and $m=3$ admits at least one optimal solution for every such $\delta$.
\end{thm}

\begin{proof}
Let $f=F_T^{-1}$, $v_T=\operatorname{Var}(T)$, and $v_S=\operatorname{Var}(S)$. Then, $f$ is continuous and strictly increasing on $(0,1)$. 
We consider three cases. 

\emph{Case 1: $v_S=v_T$.}
For any partition $Q^*$ with $0 < q_1^* < q_2^* < q_3^* = 1$, the choice $A^*=\mathbf{1}$ satisfies the required moment equations.

\emph{Case 2: $0<v_S<v_T$.}
Since $f$ is continuous and strictly increasing on $(0,1)$, there exists a unique $u_0\in(0,1)$ such that $f(u_0)=\mu$. For sufficiently small $h>0$, define
\[
M_h(a)=\frac{1}{h}\int_a^{a+h}f(u)\,du,\quad u_0-h\leq a\leq u_0
\]
which represents the conditional mean of \(T\) over the quantile interval \((a,a+h)\) of length \(h\). Since $f$ is strictly increasing, $M_h(u_0-h)<\mu<M_h(u_0).$ Moreover, $M_h$ is continuous. Hence, there exists $a_h\in(u_0-h,u_0)$ such that $M_h(a_h)=\mu.$ Since $[a_h,a_h+h]\subset[u_0-h,u_0+h]$ and $f$ is continuous at $u_0$ we find that 
\[
\begin{aligned}
0\leq v(h)
&:=\frac{1}{h}\int_{a_h}^{a_h+h}(f(u)-\mu)^2\,du \leq\sup_{|u-u_0|\leq h}|f(u)-\mu|^2
\to 0
\end{aligned}
\]
as $h\downarrow0$. Choose $h$ sufficiently small that $v(h)<v_S$, and set
\[
Q^*=(a_h,a_h+h,1)',\quad A=(0,h^{-1},0)'.
\]
By the definition of $B_{3,3}(Q^*)$ and the preceding identities,
\[
B_{3,3}(Q^*)\mathbf{1}
= (1, \mu, \mu^2+v_T)',
\quad
B_{3,3}(Q^*)A
= (1, \mu, \mu^2+v(h))'.
\]
Let $A^*=(1-\theta)\mathbf{1}+\theta A$, where $\theta=\frac{v_T-v_S}{v_T-v(h)}\in(0,1)$. Then, $A^*>0$ and $B_{3,3}(Q^*)A^* = c_3$.

\emph{Case 3: $v_T<v_S<V_{\max}$.}
For $0<\varepsilon<1/2$, consider the partition $Q^*=(\varepsilon,1-\varepsilon,1)'$. Let $\mu_1(\varepsilon)$ and $\mu_3(\varepsilon)$ denote the conditional means over $(0,\varepsilon)$ and $(1-\varepsilon,1)$, respectively, and let $v_1(\varepsilon)$ and $v_3(\varepsilon)$ denote the corresponding conditional variances. Since $f$ is strictly increasing, $\mu_1(\varepsilon)<\mu<\mu_3(\varepsilon).$ Define $A=(\varepsilon^{-1}\lambda_1(\varepsilon),0,\varepsilon^{-1}\lambda_3(\varepsilon))'$, where 
\(
\lambda_1(\varepsilon)=\frac{\mu_3(\varepsilon)-\mu}{\mu_3(\varepsilon)-\mu_1(\varepsilon)}, \lambda_3(\varepsilon)=\frac{\mu-\mu_1(\varepsilon)}{\mu_3(\varepsilon)-\mu_1(\varepsilon)}. 
\)
Clearly, $\lambda_1(\varepsilon),\lambda_3(\varepsilon)\in(0,1)$, and
\[
\lambda_1(\varepsilon)+\lambda_3(\varepsilon)=1,\quad
\lambda_1(\varepsilon)\mu_1(\varepsilon)+\lambda_3(\varepsilon)\mu_3(\varepsilon)=\mu.
\]
The step-weighted distribution associated with $Q^*$ and $A$ has mean $\mu$ and variance 
\[
v(\varepsilon)=\lambda_1(\varepsilon)v_1(\varepsilon)+\lambda_3(\varepsilon)v_3(\varepsilon)+(\mu-\mu_1(\varepsilon))(\mu_3(\varepsilon)-\mu).
\]
As $\varepsilon\downarrow0$, the conditional means $\mu_1(\varepsilon)$ and $\mu_3(\varepsilon)$ converge to $L$ and $R$, respectively, in the extended real sense. If both endpoints are finite, the conditional variances tend to zero, and hence $v(\varepsilon)\to(\mu-L)(R-\mu).$ If at least one endpoint is infinite, the last term in the expression for $v(\varepsilon)$ diverges to infinity. Therefore, in either case, $v(\varepsilon)\to V_{\max}.$ Since $v_S<V_{\max}$, we may choose $\varepsilon$ sufficiently small that $v_S < v(\varepsilon)$. By the definition of $B_{3,3}(Q^*)$ and the preceding identities,
\[
B_{3,3}(Q^*)\mathbf{1}= (1, \mu, \mu^2+v_T)', 
\quad
B_{3,3}(Q^*)A= (1, \mu, \mu^2+v(\varepsilon))'.
\]
Let $A^*=(1-\theta)\mathbf{1}+\theta A$, where $\theta=\frac{v_S-v_T}{v(\varepsilon)-v_T}\in(0,1).$ Then, $A^*>0$ and $B_{3,3}(Q^*)A^*=c_3.$ 

Therefore, in all three cases, there exist a partition $Q^*=(q_1^*,q_2^*, q_3^*)'$ with $0 < q_1^* < q_2^* < q_3^* = 1$ and a strictly positive weight vector $A^* = (a_1^*, a_2^*, a_3^*)'$ satisfying $B_{3,3}(Q^*)A^*=c_3.$ 
Define
\[
\delta_0=\min\{q_1^*,q_2^*-q_1^*,1-q_2^*\}>0.
\] 
For every $0<\delta\leq\delta_0$, we have $Q^*\in\mathcal{Q}_{3,\delta}$. By Theorem \ref{thm:moment-matrix-rank}, $B_{3,3}(Q^*)$ is invertible, and therefore $B_{3,3}(Q^*)^{-1}c_3=A^*>0.$ It follows that $Q^*\in\mathcal{Q}_{3,\delta}^{\mathrm{feas}}$ so that $\mathcal{Q}_{3,\delta}^{\mathrm{feas}}\neq\varnothing.$ The existence of an optimal solution follows from Theorem \ref{1010849}.
\end{proof}

Theorem \ref{1016884} provides a sufficient condition for the existence of an optimal solution to the three-step joint minimum-deformation problem. Specifically, under the stated conditions, if the target variance is positive and below the upper bound determined by the support of $T$, then the feasible partition set is nonempty for every sufficiently small positive $\delta$. Hence, the joint minimum-deformation problem admits an optimal solution without requiring a separate feasibility assumption.

The results of this section establish a theoretical framework for minimum deformation under exact moment-matching constraints. For a fixed quantile partition, the correction weights are determined by a strictly convex optimization problem, while allowing the partition to vary leads to a joint optimization problem. The rank structure of the moment matrix further identifies \(m=r+1\) as a natural parsimonious choice, under which moment matching uniquely determines the weights and minimum deformation selects the optimal quantile partition. In the case of matching the first two moments, we also establish a sufficient condition for the existence of a feasible three-step correction and hence an optimal solution. In the next section, we apply these theoretical results to the step-weighted correction of comonotonic approximations for sums of dependent lognormal random variables.

\section{Applications to sums of lognormals}\label{sec:numerical}
In this section, we apply the minimum-deformation framework developed in Section \ref{sec:min-deformation} to the approximation of sums of dependent lognormal random variables. We consider three starting comonotonic approximations: the comonotonic upper bound, the Taylor-based (TB) approximation, and the maximal-variance (MV) approximation. Starting from each of these approximations, we construct step-weighted corrections that match the prescribed moments of the original sum. Particular attention is paid to the choice \(m=r+1\), for which the correction weights are uniquely determined by the moment-matching equations once the quantile partition is fixed. The remaining task is therefore to select, among the feasible quantile partitions, one that minimizes the Pearson deformation.

\subsection{Starting comonotonic approximations} \label{1017586}
In this subsection, we briefly recall three comonotonic approximations for sums of lognormal random variables. Consider the random sum
\begin{equation} \label{1023589}
S=\sum_{i=1}^{n}\alpha_i e^{Z_i}
\end{equation}
where $\alpha_1,\ldots,\alpha_n$ are positive real numbers and $(Z_1,\ldots,Z_n)'$ follows a multivariate normal distribution with mean vector $\mu$ and covariance matrix $\Sigma$. Throughout this section, we assume that the covariance matrix $\Sigma$ is positive definite and satisfies  
\begin{equation}\label{1023593}
\Sigma_{ij}\geq0, \,\,  i,j=1,\ldots,n.
\end{equation}
In particular, this setting includes the Brownian-motion covariance structure
\begin{equation} \label{1023597}
\Sigma_{ij}=\sigma^2\min(t_i,t_j), \,\, i,j=1,\ldots,n 
\end{equation}
with $0<t_1<\cdots<t_n$ and $\sigma>0$, which is widely used in financial modeling.

Following \cite{Dhaene et al. (2002a), Dhaene et al. (2002b), DVG2005, KDG2000, VCD2008}, we consider the convex-order upper and lower bounds $S^c$ and $S^l$ as approximations of $S$. The upper bound is given by
\begin{equation} \label{1023605}
S^c=\sum_{i=1}^n\alpha_i e^{\mathrm{E}[Z_i]+\sigma_{Z_i}\Phi^{-1}(U)}
\end{equation}
where $U$ is uniformly distributed on $(0,1)$. For the lower bound, we consider the conditioning random variable
\begin{equation} \label{1023609}
\Lambda=\sum_{i=1}^n\lambda_i Z_i
\end{equation}
where \(\lambda_1,\ldots,\lambda_n\) are real numbers, not all zero. Denote the mean and variance of $\Lambda$ by $\mathrm{E}[\Lambda]$ and $\sigma_\Lambda^2$, respectively. Then the lower bound $S^l$ can be expressed as
\begin{equation} \label{1023613}
S^l=\sum_{i=1}^n\alpha_i\exp\left(\mathrm{E}[Z_i]+\frac12(1-r_i^2)\sigma_{Z_i}^2+r_i\sigma_{Z_i}\Phi^{-1}(V)\right)
\end{equation}
where $V=\Phi((\Lambda-\mathrm{E}[\Lambda])/\sigma_\Lambda)$ is uniformly distributed on $(0,1)$, and $r_i$ denotes the correlation coefficient between $Z_i$ and $\Lambda$. If $r_i\geq0$ for all $i$, each term in $S^l$ is a non-decreasing function of the common random variable $V$. Hence, by Lemma \ref{626115}, $S^l$ is a comonotonic sum. Consequently, the quantiles and conditional tail expectations of $S^l$ can be obtained by summing those of its comonotonic components.

The approximation performance of the lower bound $S^l$ depends on the choice of the conditioning random variable $\Lambda$. We consider two representative approaches: the Taylor-based (TB) approximation $S^l_{TB}$ and the maximal-variance (MV) approximation $S^l_{MV}$. Following \cite{Dhaene et al. (2002a), Dhaene et al. (2002b), KDG2000}, the TB approximation $S^l_{TB}$ is obtained by choosing $\lambda_i=\alpha_i\exp(\mathbb{E}[Z_i])$, while the MV approximation $S^l_{MV}$, following \cite{VCD2008, VDG2005a, VHD2005b}, is obtained by choosing $\lambda_i=\alpha_i\mathbb{E}[e^{Z_i}]$, for $i=1,\ldots,n$. Under assumption (\ref{1023593}), both choices yield nonnegative correlation coefficients $r_i$, so that the resulting lower bounds are comonotonic sums. Although all three approximations $S^c$, $S^l_{TB}$, and $S^l_{MV}$ are tractable, their approximation performance may vary with the model parameters and the risk measures considered. All three approximations preserve the mean of $S$. However, the variance of $S^c$ is no smaller than that of $S$, whereas the variances of $S^l_{TB}$ and $S^l_{MV}$ are no greater than that of $S$. This motivates the application of the minimum-deformation framework developed in Section \ref{sec:min-deformation} to achieve exact moment matching while preserving comonotonicity.

\subsection{Minimum-deformation comonotonic approximations} \label{1024659}
We now apply the minimum-deformation framework developed in Section \ref{sec:min-deformation} to the three comonotonic approximations \(S^c\), \(S^l_{TB}\), and \(S^l_{MV}\). Throughout this subsection, we adopt the canonical choice \(m=r+1\), as motivated in Subsection \ref{subsec:step-selection}, with \(r\geq 1\) denoting the highest raw-moment order to be matched. 

Let $\mu_i=\mathbb{E}[Z_i]$ and $\Sigma_{ij}=\operatorname{Cov}(Z_i,Z_j)$. For each starting approximation $T\in\{S^c,S^l_{TB},S^l_{MV}\}$, define
\[
\beta_i=
\begin{cases}
\alpha_i\exp(\mu_i), & T=S^c,\\
\alpha_i\exp\left(\mu_i+\dfrac{1-r_i^2}{2}\Sigma_{ii}\right), & T=S^l_{TB}\text{ or }S^l_{MV},
\end{cases}
\]
and
\[
\gamma_i=
\begin{cases}
\sqrt{\Sigma_{ii}}, & T=S^c,\\
r_i\sqrt{\Sigma_{ii}}, & T=S^l_{TB}\text{ or }S^l_{MV},
\end{cases}
\quad i=1,\ldots,n,
\]
with $r_i=\operatorname{Corr}(Z_i,\Lambda)$ for the corresponding TB or MV approximation. Under assumption (\ref{1023593}), we have $\beta_i>0$ and $\gamma_i>0$ for all $i$. Consequently, the quantile function of $T$ admits the representation
\begin{equation} \label{1024667}
F_T^{-1}(u)=\sum_{i=1}^n\beta_i\exp\left(\gamma_i\Phi^{-1}(u)\right),\quad 0<u<1.
\end{equation}
It follows that \(F_T^{-1}\) is continuous and strictly increasing on \((0,1)\). Moreover, both \(S\) and \(T\) have finite moments of every positive integer order. 

Let \(c_{r+1}=(1,\mathbb E[S],\ldots,\mathbb E[S^r])'\) be the target moment vector. We first derive an explicit representation for the moment matrix $B_{r+1, r + 1}(Q)$ associated with the starting approximation $T$. To this end, for $t\in\mathbb R$ and $0\leq a<b\leq1$, define
\[
H(t;a,b)
=\exp({t^2}/{2})[\Phi(\Phi^{-1}(b)-t)-\Phi(\Phi^{-1}(a)-t)]
\]
with $\Phi^{-1}(0)=-\infty$ and $\Phi^{-1}(1)=+\infty$. In the following, for any integer $r\geq1$, a numerical vector $Q=(q_1,\ldots,q_{r+1})'$ is called a partition of the unit interval $[0,1]$ if
\[
0=q_0<q_1<\cdots<q_r<q_{r+1}=1.
\]

\begin{lem} \label{1023687}
Consider the sum $S$ of lognormal random variables defined in (\ref{1023589}), and let $T$ be any one of the three comonotonic approximations $S^c$, $S^l_{TB}$, and $S^l_{MV}$. For any integer $r\geq1$, let $Q=(q_1,\ldots,q_{r+1})'$ be a partition of the unit interval. Then the moment matrix $B_{r+1,r+1}(Q)$ associated with $T$ is nonsingular, with entries given by
$$M_{0j}(Q)=q_j-q_{j-1}, \,\, j=1,\ldots,r+1,$$ and
\[
M_{kj}(Q)
=
\sum_{i_1=1}^{n}\cdots\sum_{i_k=1}^{n}
\left(\prod_{\ell=1}^{k}\beta_{i_\ell}\right)
H\left(
\sum_{\ell=1}^{k}\gamma_{i_\ell};
q_{j-1},q_j
\right), \,\, k=1,\ldots,r, j=1,\ldots,r+1. 
\]
\end{lem}
\begin{proof}
For any $t\in\mathbb R$ and $0\leq a<b\leq1$, 
$$
H(t;a,b) = \int_a^b \exp\left(t\Phi^{-1}(u)\right)du.
$$ 
For $k=1,\ldots,r$, the multinomial expansion in its indexed form yields
\[
\left(F_T^{-1}(u)\right)^k
=
\sum_{i_1=1}^{n}\cdots\sum_{i_k=1}^{n}
\left(\prod_{\ell=1}^{k}\beta_{i_\ell}\right)
\exp\left[
\left(\sum_{\ell=1}^{k}\gamma_{i_\ell}\right)
\Phi^{-1}(u)
\right].
\]
Integrating over $(q_{j-1},q_j)$ gives the stated expression for $M_{kj}(Q)$. The expression for $M_{0j}(Q)$ follows immediately from its definition. Since $T$ has a continuous distribution function and finite moments, Theorem \ref{thm:moment-matrix-rank} implies that $B_{r+1,r+1}(Q)$ is nonsingular.
\end{proof}

The preceding result provides an explicit representation of the moment matrix in terms of finite sums, the standard normal distribution function, and its quantile function, without requiring numerical integration or inversion of the distribution function of $T$. We now consider the minimum-deformation comonotonic approximation of \(S\) for a fixed quantile partition. 

\begin{thm} \label{1023722}
Consider the sum $S$ of lognormal random variables defined in (\ref{1023589}), and let $T$ be any one of the three comonotonic approximations $S^c$, $S^l_{TB}$, and $S^l_{MV}$. For any integer $r\geq 1$, let $Q=(q_1,\ldots,q_{r+1})'$ be a partition of the unit interval and define $A(Q)=B_{r+1,r+1}(Q)^{-1}c_{r+1}.$ If $A(Q)\geq 0$, then $T_{A(Q),Q}$ is the unique minimum-deformation comonotonic approximation of $S$ among all step-weighted versions of $T$ based on $Q$ that match the first $r$ moments of $S$.
\end{thm}

\begin{proof}
By Lemma \ref{1023687}, the moment matrix $B_{r+1,r+1}(Q)$ is nonsingular. Hence, the moment-matching equations $B_{r+1,r+1}(Q)A=c_{r+1}$ admit the unique solution
\[
A(Q)=B_{r+1,r+1}(Q)^{-1}c_{r+1}.
\]
If $A(Q)\geq 0$, the feasible set of the fixed-partition minimum-deformation problem consists of the single vector $A(Q)$. Hence, $T_{A(Q),Q}$ is the unique minimum-deformation comonotonic approximation of $S$ among all step-weighted versions of $T$ based on $Q$ that match the first $r$ moments of $S$.
\end{proof}

From Theorem \ref{1023722} we see that, for a fixed quantile partition $Q$ with $m=r+1$, the moment-matching conditions uniquely determine the correction weights. Whenever these weights are nonnegative, they yield the minimum-deformation comonotonic approximation of $S$ within the class of step-weighted versions of $T$ based on $Q$ that match the first $r$ moments of $S$. The remaining question is how to choose the quantile partition to minimize the deformation while preserving moment matching. We therefore turn to the joint minimum-deformation comonotonic approximation problem, in which the quantile partition is also allowed to vary.

\begin{thm} \label{1024736}
Consider the sum $S$ of lognormal random variables defined in (\ref{1023589}), and let $T$ be any one of the three comonotonic approximations $S^c$, $S^l_{TB}$, and $S^l_{MV}$. For an integer $r\geq 1$, suppose that $\mathcal{Q}_{r+1,\delta}^{\mathrm{feas}}\neq\varnothing$ for some $0<\delta\leq 1/(r+1)$. Then there exists a partition $Q^* \in \mathcal Q_{r+1,\delta}^{\mathrm{feas}}$ such that $T_{A^*,Q^*}$ with $A^*=B_{r+1,r+1}(Q^*)^{-1}c_{r+1}$ defines a minimum-deformation comonotonic approximation of $S$ matching the first $r$ moments of $S$ within the class of step-weighted versions of $T$ with partitions in $\mathcal{Q}^{\mathrm{feas}}_{r+1,\delta}$. 
\end{thm}

\begin{proof}
Since \(S\) is a finite sum of lognormal random variables and \(T\) has the representation given in (\ref{1024667}), they both have finite raw moments of every positive integer order. Thus, the moment conditions required in Theorem \ref{1010849} are satisfied for every integer $r\geq 1$. By Lemma \ref{1023687}, the matrix $B_{r+1,r+1}(Q)$ is nonsingular for every admissible partition $Q$. Hence, the moment-matching equations uniquely determine the weight vector $A(Q)=B_{r+1,r+1}(Q)^{-1}c_{r+1}$. Since $\mathcal{Q}_{r+1,\delta}^{\mathrm{feas}}\neq\varnothing$, Theorem \ref{1010849} guarantees the existence of an optimal partition $Q^*\in\mathcal{Q}_{r+1,\delta}^{\mathrm{feas}}$ minimizing the Pearson deformation. The corresponding weight vector $A^*=B_{r+1,r+1}(Q^*)^{-1}c_{r+1}$ is nonnegative and satisfies the prescribed moment-matching conditions. Furthermore, by Theorem \ref{1024148}, the step-weighted correction preserves the comonotonic sum structure. Consequently, $T_{A^*,Q^*}$ is a minimum-deformation comonotonic approximation of $S$ matching its first $r$ moments within the specified class.
\end{proof}

Theorem \ref{1024736} establishes the existence of a minimum-deformation comonotonic approximation of \(S\) for each of the three starting approximations \(S^c\), \(S^l_{TB}\), and \(S^l_{MV}\), whenever the corresponding feasible partition set is nonempty. Under the choice \(m=r+1\), the moment-matching conditions uniquely determine the correction weights for each fixed partition, while the minimum-deformation principle selects an optimal partition among all feasible candidates. However, Theorem \ref{1024736} does not guarantee that the feasible partition set is nonempty. The following theorem addresses this issue for three-step corrections by establishing, under suitable conditions, the existence of feasible partitions and hence of minimum-deformation comonotonic approximations matching the first two moments of \(S\).

\begin{thm} \label{1024746}
Consider the sum $S$ of lognormal random variables defined in (\ref{1023589}), and let $T$ be any one of the three comonotonic approximations $S^c$, $S^l_{TB}$, and $S^l_{MV}$. Then there exists $\delta_0>0$ such that, for every $0<\delta\leq\delta_0$, the feasible partition set $\mathcal{Q}_{3,\delta}^{\mathrm{feas}}$ is nonempty. Consequently, there exists a partition $Q^*\in\mathcal{Q}_{3,\delta}^{\mathrm{feas}}$ such that $T_{A^*,Q^*}$ with $A^*=B_{3,3}(Q^*)^{-1}c_3$ defines a minimum-deformation comonotonic approximation of $S$ matching its first two moments within the class of three-step-weighted versions of $T$ with partitions in $\mathcal{Q}_{3,\delta}^{\mathrm{feas}}$. Moreover, $\delta_0$ may be chosen uniformly for the three comonotonic approximations $S^c$, $S^l_{TB}$, and $S^l_{MV}$.
\end{thm}

\begin{proof}
First, we note that $E[T]=E[S]$. Moreover, by (\ref{1024667}), $T$ admits a representation of the form
\(
T\stackrel{d}{=}\sum_{i=1}^{n}\beta_i e^{\gamma_i Z},
\)
where $Z$ is standard normal and $\beta_i>0$, $\gamma_i>0$ for all $i=1,\ldots,n$. Consequently, the distribution function of $T$ is continuous on $\mathbb{R}$ and strictly increasing on $(0,\infty)$, with $\operatorname{ess\,inf}T=0$ and $\operatorname{ess\,sup}T=+\infty$. Since $S$ is a nondegenerate finite sum of lognormal random variables, $0<\operatorname{Var}(S)<\infty$. Thus, all the conditions of Theorem \ref{1016884} are satisfied, and there exists $\delta_T>0$ such that $\mathcal{Q}_{3,\delta}^{\mathrm{feas}}\neq\varnothing$ for every $0<\delta\leq\delta_T$. Applying this argument to the three approximations separately and taking $\delta_0$ to be the minimum of the three corresponding positive constants, we obtain a common $\delta_0>0$ for which $\mathcal{Q}_{3,\delta}^{\mathrm{feas}}$ is nonempty for all three approximations whenever $0<\delta\leq\delta_0$. By Theorem \ref{1024736} with $r=2$, there exists an optimal partition $Q^*\in\mathcal{Q}_{3,\delta}^{\mathrm{feas}}$ such that $T_{A^*,Q^*}$, with $A^*=B_{3,3}(Q^*)^{-1}c_3$, defines a minimum-deformation comonotonic approximation of $S$ matching its first two moments within the specified class.
\end{proof}

Theorem \ref{1024746} strengthens Theorem \ref{1024736} in the three-step case by guaranteeing the nonemptiness of the feasible partition set for sufficiently small positive values of $\delta$. In particular, for each of the three starting approximations $S^c$, $S^l_{TB}$, and $S^l_{MV}$, a minimum-deformation comonotonic approximation matching the first two moments of $S$ exists under the assumptions of Subsection \ref{1017586} whenever $\delta$ is sufficiently small. Moreover, the threshold $\delta_0$ can be chosen uniformly across the three methods. Together with Theorems \ref{1023722} and \ref{1024736}, this result establishes the theoretical foundation for minimum-deformation comonotonic approximations of sums of lognormal random variables. The following section presents numerical illustrations of these theoretical results and examines the performance of the resulting minimum-deformation comonotonic approximations.

\section{Numerical illustrations} \label{1031742}
In this section, we return to the sum of lognormal random variables as described in Section \ref{sec:intro} and illustrate the minimum-deformation approach developed in the preceding sections by using a classical numerical example. This example has been considered extensively in the literature, see \cite{Dhaene et al. (2002b), DVG2005, VCD2008, VHD2005b} for instance. 
\subsection{Model description} \label{1030758}
In this subsection, we describe the model and fix the notations which will be used in the following subsections. We consider the sum
\[
S=\sum_{i=1}^{n}\alpha_i e^{Z_i}
\]
where \(\alpha_1,\ldots,\alpha_n\) are positive real numbers and \((Z_1,\ldots,Z_n)\) is a random vector following a multivariate normal distribution with
\[
\mathbb E[Z_i]=-i\left(\mu-\frac12\sigma^2\right), \,\, i=1,\ldots,n
\]
and
\[
\operatorname{Cov}(Z_i,Z_j)=\min(i,j)\sigma^2, \,\,  i,j=1,\ldots,n, \sigma>0.
\]

As in our previous paper \cite{HuangStep}, we consider the case where \(n=20\), \(\mu=0.075\), \(\sigma=0.15\), and \(\alpha_i=1\) for \(i=1,\ldots,n\). In Subsections \ref{1030775}–\ref{116802}, we fix the moment-matching order at \(r=2\) and the number of steps at \(m=3\). Higher-order moment matching is considered separately in Subsection \ref{116870}. Unlike in \cite{HuangStep}, however, the quantile partition \(Q\) is not fixed in advance. In \cite{HuangStep}, we chose the equally spaced partition \(Q=(1/3,2/3,1)'\) and then determined the correction weights \(A\) by solving the moment-matching equations. In the present paper, the partition \(Q\) and the correction weights \(A\) are jointly selected according to the minimum-deformation principle developed in the preceding sections. In Subsections \ref{1030775}–\ref{116802}, we set \(\delta=0.001\) for all three starting comonotonic approximations. If the minimum interval-width constraint is active at an optimal solution, one may choose a smaller value of $\delta$ to enlarge the feasible set.

Throughout this section, we use quantiles and conditional tail expectations (CTE's) as criteria to assess the approximation performance of the different methods. The corresponding values are compared with the Monte Carlo results as reported in \cite{HuangStep}, based on 500,000 pairs of antithetic random paths (1,000,000 simulated paths in total). The relative deviations are expressed as percentages of the corresponding Monte Carlo estimates, and the mean absolute deviation (MAD) is the arithmetic mean of their absolute values. The smallest absolute relative deviations and MAD's are highlighted in bold in the tables.

\subsection{Minimum-deformation approximations based on \(S^c\)} \label{1030775}
In this subsection, the starting comonotonic approximation of $S$ is the comonotonic sum 
\begin{equation*} 
S^c = \sum^n_{i = 1}\alpha_ie^{E[Z_i] + \sigma_{Z_i}\Phi^{-1}(U)}
\end{equation*}
where $U$ is a random variable uniformly distributed on the unit interval $(0, 1)$. 

As shown in \cite{HuangStep}, the equally spaced partition $Q=(1/3,2/3,1)'$ and the corresponding weights $A=(0.5284514,1.7275870,0.7439616)'$ satisfy the moment-matching conditions, with a Pearson deformation of $0.27243218$. For the minimum-deformation correction, numerical optimization yields $Q^*=(0.2822814,0.9817161,1)'$ and $A^*=(0.8163659,1.1002529,0)'$, with a Pearson deformation of $0.03483262$. Compared with the equally spaced correction, the Pearson deformation is reduced by approximately $87.21\%$.

Tables~\ref{tab:Sc_quantiles} and \ref{tab:Sc_CTE} show the relative performance of the three approximations. For the probability levels 0.25,0.50,0.75,0.90,0.95, the minimum-deformation correction $(S^c)_{\omega^*}$ achieves the smallest MADs for both quantiles and CTEs, reducing them to $1.83\%$ and $1.79\%$, respectively. For the upper probability levels 0.95,0.975,0.99,0.995,0.999, however, the equally spaced correction $(S^c)_{\omega}$ performs better, with MADs of $6.34\%$ for quantiles and $9.03\%$ for CTEs, compared with $9.88\%$ and $17.89\%$ for $(S^c)_{\omega^*}$. In particular, at $p=0.999$, the minimum-deformation correction $(S^c)_{\omega^*}$ produces relative deviations of $-27.40\%$ and $-36.11\%$ for the quantile and CTE, respectively, indicating substantial underestimation of the extreme upper tail. This behavior is closely related to the zero weight assigned to the last interval in the numerically obtained minimum-deformation solution, which truncates the upper tail of the corrected distribution.

\begin{table}[htbp]
\centering
\begin{tabular}{lrrrrrr}
\toprule
Method & $p=0.25$ & $p=0.5$ & $p=0.75$ & $p=0.9$ & $p=0.95$ & MAD \\
\midrule
$S^c$ & -6.30\% & -2.26\% & \textbf{+1.87\%} & +5.63\% & +7.86\% & 4.78\% \\
$(S^c)_{\omega}$ & +8.04\% & \textbf{-0.15\%} & -8.19\% & \textbf{-2.15\%} & \textbf{+1.01\%} & 3.91\% \\
$(S^c)_{\omega^*}$ & \textbf{-0.58\%} & +0.52\% & +2.48\% & +3.32\% & +2.24\% & \textbf{1.83\%} \\
MC & 8.9167 & 11.1994 & 14.2277 & 17.8198 & 20.4882 & NA \\
\midrule
Method & $p=0.95$ & $p=0.975$ & $p=0.99$ & $p=0.995$ & $p=0.999$ & MAD \\
\midrule
$S^c$ & +7.86\% & +9.88\% & +12.26\% & +13.98\% & +16.71\% & 12.14\% \\
$(S^c)_{\omega}$ & \textbf{+1.01\%} & +3.64\% & \textbf{+6.58\%} & \textbf{+8.60\%} & \textbf{+11.88\%} & \textbf{6.34\%} \\
$(S^c)_{\omega^*}$ & +2.24\% & \textbf{-0.52\%} & -6.69\% & -12.58\% & -27.40\% & 9.88\% \\
MC & 20.4882 & 23.1697 & 26.8056 & 29.6348 & 36.8715 & NA \\
\bottomrule
\end{tabular}
\caption{Approximations for selected quantiles of $S$ based on $S^c$ with $\mu=0.075$, $\sigma=0.15$, $n=20$, and yearly payments of 1.}
\label{tab:Sc_quantiles}
\end{table}

\begin{table}[htbp]
\centering
\begin{tabular}{lrrrrrr}
\toprule
Method & $p=0.25$ & $p=0.5$ & $p=0.75$ & $p=0.9$ & $p=0.95$ & MAD \\
\midrule
$S^c$ & +1.69\% & +3.59\% & +6.23\% & +9.06\% & +10.92\% & 6.30\% \\
$(S^c)_{\omega}$ & -1.10\% & -2.59\% & -1.60\% & +2.45\% & +4.87\% & 2.52\% \\
$(S^c)_{\omega^*}$ & \textbf{+0.85\%} & \textbf{+1.17\%} & \textbf{+0.95\%} & \textbf{-1.32\%} & \textbf{-4.64\%} & \textbf{1.79\%} \\
MC & 13.5965 & 15.3713 & 18.1574 & 21.7366 & 24.4722 & NA \\
\midrule
Method & $p=0.95$ & $p=0.975$ & $p=0.99$ & $p=0.995$ & $p=0.999$ & MAD \\
\midrule
$S^c$ & +10.92\% & +12.57\% & +14.56\% & +15.77\% & +17.82\% & 14.33\% \\
$(S^c)_{\omega}$ & +4.87\% & \textbf{+6.93\%} & \textbf{+9.31\%} & \textbf{+10.77\%} & \textbf{+13.26\%} & \textbf{9.03\%} \\
$(S^c)_{\omega^*}$ & \textbf{-4.64\%} & -9.26\% & -16.68\% & -22.75\% & -36.11\% & 17.89\% \\
MC & 24.4722 & 27.2812 & 31.1339 & 34.2242 & 42.0901 & NA \\
\bottomrule
\end{tabular}
\caption{Approximations for selected CTE’s of $S$ based on $S^c$ with $\mu=0.075$, $\sigma=0.15$, $n=20$, and yearly payments of 1.}
\label{tab:Sc_CTE}
\end{table}

These numerical results reveal a fundamental feature of our approach. The objective is not to minimize the approximation error relative to the true distribution of $S$, but to preserve the distribution of the starting comonotonic approximation as much as possible while satisfying the prescribed moment-matching constraints. Consequently, a small Pearson deformation does not necessarily imply an accurate approximation to the distribution of $S$, nor does matching a finite number of moments guarantee improved accuracy of quantiles or CTE's. 

The quality of the starting comonotonic approximation is therefore crucial to the effectiveness of the proposed method. If the starting approximation is already sufficiently accurate, the minimum-deformation principle allows moment matching while limiting changes to its distributional characteristics. Conversely, if the starting approximation is poor, minimizing its deformation cannot be expected to eliminate the existing approximation errors and may even adversely affect the upper tail. Thus, the minimum-deformation correction is primarily intended to refine a good starting approximation rather than to compensate for a poor one. This consideration motivates the following numerical investigations based on the TB and MV comonotonic approximations.

\subsection{Minimum-deformation approximations based on \(S^l_{\mathrm{TB}}\)} 
In this subsection, the starting comonotonic approximation of $S$ is the Taylor-based (TB) lower bound 
\[
S_{\mathrm{TB}}^l
=
\sum_{i=1}^{n}
\alpha_i
\exp\left(
\mathbb{E}[Z_i]
+\frac{1}{2}(1-r_{i,\mathrm{TB}}^2)\sigma_{Z_i}^2
+r_{i,\mathrm{TB}}\sigma_{Z_i}\Phi^{-1}(U)
\right)
\]
where $U$ is uniformly distributed on $(0,1)$ and $r_{i,\mathrm{TB}}=\operatorname{Corr}(Z_i,\Lambda_{\mathrm{TB}})$.

As shown in \cite{HuangStep}, the equally spaced partition $Q=(1/3,2/3,1)'$ and the corresponding weights $A=(1.01309659,0.97932878,1.00757463)'$ satisfy the moment-matching conditions, with a Pearson deformation of $0.00021873$. For the minimum-deformation correction, numerical optimization yields $Q^*=(0.2461594,0.9920129,1)'$ and $A^*=(1.0051506,0.9977558,1.0508258)'$, with a Pearson deformation of $0.00003092$. Compared with the equally spaced correction, the Pearson deformation is reduced by approximately $85.86\%$.

Tables \ref{tab:TB_quantiles} and \ref{tab:TB_CTE} show the relative performance of the three approximations. For the probability levels 0.25,0.50,0.75,0.90,0.95, the equally spaced correction $(S_{\mathrm{TB}}^l)_\omega$ reduces the CTE MAD from $0.15\%$ to $0.09\%$, but increases the quantile MAD from $0.06\%$ to $0.14\%$. For the upper probability levels 0.95,0.975,0.99,0.995,0.999, the equally spaced correction improves both quantile and CTE approximations, reducing their MAD's from $0.42\%$ and $0.89\%$ to ${0.30\%}$ and $0.79\%$, respectively. In contrast, the minimum-deformation correction $(S_{\mathrm{TB}}^l)_{\omega^*}$ yields the smallest MAD's in both probability groups, with $0.04\%$ and $0.25\%$ for quantiles and $0.02\%$ and $0.39\%$ for CTE's, respectively. These results indicate that the minimum-deformation correction preserves the high accuracy of the TB approximation while achieving exact moment matching, with particularly noticeable improvements in CTE approximation.

\begin{table}[htbp]
\centering
\begin{tabular}{lrrrrrr}
\hline
Method & $p=0.25$ & $p=0.5$ & $p=0.75$ & $p=0.9$ & $p=0.95$ & MAD \\
\hline
$S_{\mathrm{TB}}^l$ & $\mathbf{+0.04\%}$ & $\mathbf{+0.03\%}$ & $+0.01\%$ & $-0.05\%$ & $-0.16\%$ & $0.06\%$ \\
$(S_{\mathrm{TB}}^l)_\omega$ & $-0.29\%$ & $-0.05\%$ & $+0.22\%$ & $+0.11\%$ & $\mathbf{-0.02\%}$ & $0.14\%$ \\
$(S_{\mathrm{TB}}^l)_{\omega^*}$ & $-0.09\%$ & $-0.03\%$ & $\mathbf{-0.01\%}$ & $\mathbf{-0.01\%}$ & $-0.05\%$ & $\mathbf{0.04\%}$ \\
MC & $8.9167$ & $11.1994$ & $14.2277$ & $17.8198$ & $20.4882$ & NA \\
\hline
Method & $p=0.95$ & $p=0.975$ & $p=0.99$ & $p=0.995$ & $p=0.999$ & MAD \\
\hline
$S_{\mathrm{TB}}^l$ & $-0.16\%$ & $-0.22\%$ & $-0.30\%$ & $-0.31\%$ & $-1.10\%$ & $0.42\%$ \\
$(S_{\mathrm{TB}}^l)_\omega$ & $\mathbf{-0.02\%}$ & $-0.10\%$ & $\mathbf{-0.19\%}$ & $\mathbf{-0.21\%}$ & $-1.01\%$ & ${0.30\%}$ \\
$(S_{\mathrm{TB}}^l)_{\omega^*}$ & $-0.05\%$ & $\mathbf{+0.03\%}$ & $+0.31\%$ & $+0.38\%$ & $\mathbf{-0.49\%}$ & $\mathbf{0.25\%}$ \\
MC & $20.4882$ & $23.1697$ & $26.8056$ & $29.6348$ & $36.8715$ & NA \\
\hline
\end{tabular}
\caption{Approximations for selected quantiles of $S$ based on $S_{\mathrm{TB}}^l$ with $\mu=0.075$, $\sigma=0.15$, $n=20$, and yearly payments of 1.}
\label{tab:TB_quantiles}
\end{table}

\begin{table}[htbp]
\centering
\begin{tabular}{lrrrrrr}
\hline
Method & $p=0.25$ & $p=0.5$ & $p=0.75$ & $p=0.9$ & $p=0.95$ & MAD \\
\hline
$S_{\mathrm{TB}}^l$ & $-0.04\%$ & $-0.07\%$ & $-0.12\%$ & $-0.23\%$ & $-0.32\%$ & $0.15\%$ \\
$(S_{\mathrm{TB}}^l)_\omega$ & $\mathbf{+0.01\%}$ & $+0.09\%$ & $+0.04\%$ & $-0.09\%$ & $-0.20\%$ & $0.09\%$ \\
$(S_{\mathrm{TB}}^l)_{\omega^*}$ & $-0.02\%$ & $\mathbf{-0.01\%}$ & $\mathbf{-0.00\%}$ & $\mathbf{+0.01\%}$ & $\mathbf{+0.04\%}$ & $\mathbf{0.02\%}$ \\
MC & $13.5965$ & $15.3713$ & $18.1574$ & $21.7366$ & $24.4722$ & NA \\
\hline
Method & $p=0.95$ & $p=0.975$ & $p=0.99$ & $p=0.995$ & $p=0.999$ & MAD \\
\hline
$S_{\mathrm{TB}}^l$ & $-0.32\%$ & $-0.46\%$ & $-0.65\%$ & $-0.95\%$ & $-2.08\%$ & $0.89\%$ \\
$(S_{\mathrm{TB}}^l)_\omega$ & $-0.20\%$ & $-0.35\%$ & $-0.55\%$ & $-0.86\%$ & $-1.99\%$ & $0.79\%$ \\
$(S_{\mathrm{TB}}^l)_{\omega^*}$ & $\mathbf{+0.04\%}$ & $\mathbf{+0.05\%}$ & $\mathbf{+0.01\%}$ & $\mathbf{-0.32\%}$ & $\mathbf{-1.52\%}$ & $\mathbf{0.39\%}$ \\
MC & $24.4722$ & $27.2812$ & $31.1339$ & $34.2242$ & $42.0901$ & NA \\
\hline
\end{tabular}
\caption{Approximations for selected CTE’s of $S$ based on $S_{\mathrm{TB}}^l$ with $\mu=0.075$, $\sigma=0.15$, $n=20$, and yearly payments of 1.}
\label{tab:TB_CTE}
\end{table}

These findings complement the observations in Subsection \ref{1030775}. When the starting comonotonic approximation is already highly accurate, as in the present TB example, the minimum-deformation correction can achieve moment matching with only a small change to the starting distribution. In this case, the correction also improves the overall accuracy of both quantiles and CTE's, although such improvements are not guaranteed by the minimum-deformation principle itself. We next examine whether similar behavior occurs for the MV approximation.
\subsection{Minimum-deformation approximations based on \(S^l_{\mathrm{MV}}\)} \label{116802}
In this subsection, the starting comonotonic approximation is the maximal variance (MV) lower bound 
\[
S_{\mathrm{MV}}^l
=
\sum_{i=1}^{n}
\alpha_i
\exp\left(
\mathbb{E}[Z_i]
+\frac{1}{2}(1-r_{i,\mathrm{MV}}^2)\sigma_{Z_i}^2
+r_{i,\mathrm{MV}}\sigma_{Z_i}\Phi^{-1}(U)
\right)
\]
where $U$ is uniformly distributed on $(0,1)$ and $r_{i,\mathrm{MV}}=\operatorname{Corr}(Z_i,\Lambda_{\mathrm{MV}})$.

As shown in \cite{HuangStep}, the equally spaced partition $Q=(1/3,2/3,1)'$ and the corresponding weights $A=(1.0101099,0.9840636,1.0058264)'$ satisfy the moment-matching conditions, with a Pearson deformation of $0.00013004$. For the minimum-deformation correction, numerical optimization yields $Q^*=(0.2477227,0.9921957,1)'$ and $A^*=(1.0039190,0.9982825,1.0394390)'$, with a Pearson deformation of $0.00001814$. Compared with the equally spaced correction, the Pearson deformation is reduced by approximately $86.05\%$.

Tables \ref{tab:MV_quantiles} and \ref{tab:MV_CTE} show the relative performance of the three approximations. For the probability levels 0.25,0.50,0.75,0.90,0.95, the equally spaced correction $(S_{\mathrm{MV}}^l)_\omega$ reduces the CTE MAD from $0.11\%$ to $0.05\%$, but increases the quantile MAD from $0.05\%$ to ${0.11\%}$. For the upper probability levels 0.95,0.975,0.99,0.995,0.999, the equally spaced correction improves both quantile and CTE approximations, reducing their MAD's from $0.25\%$ and $0.64\%$ to $0.16\%$ and $0.56\%$, respectively. By comparison, the minimum-deformation correction $(S_{\mathrm{MV}}^l)_{\omega^*}$ yields the smallest MAD's for the first group of quantiles and both groups of CTE's, with $0.04\%$, $0.04\%$, and $0.34\%$, respectively. For the upper quantiles, however, the equally spaced correction performs better, with a MAD of $0.16\%$, compared with $0.23\%$ for the minimum-deformation correction. These results indicate that the minimum-deformation correction preserves the high accuracy of the MV approximation while achieving exact moment matching, with particularly noticeable improvements in CTE approximation.

\begin{table}[htbp]
\centering
\begin{tabular}{lrrrrrr}
\hline
Method & $p=0.25$ & $p=0.5$ & $p=0.75$ & $p=0.9$ & $p=0.95$ & MAD \\
\hline
$S_{\mathrm{MV}}^l$ & $\mathbf{+0.03\%}$ & $\mathbf{-0.02\%}$ & $\mathbf{-0.03\%}$ & $-0.05\%$ & $-0.12\%$ & $0.05\%$ \\
$(S_{\mathrm{MV}}^l)_\omega$ & $-0.22\%$ & $-0.08\%$ & $+0.13\%$ & $+0.08\%$ & $\mathbf{-0.01\%}$ & ${0.11\%}$ \\
$(S_{\mathrm{MV}}^l)_{\omega^*}$ & $-0.07\%$ & $-0.07\%$ & $-0.05\%$ & $\mathbf{-0.02\%}$ & $-0.03\%$ & $\mathbf{0.04\%}$ \\
MC & $8.9167$ & $11.1994$ & $14.2277$ & $17.8198$ & $20.4882$ & NA \\
\hline
Method & $p=0.95$ & $p=0.975$ & $p=0.99$ & $p=0.995$ & $p=0.999$ & MAD \\
\hline
$S_{\mathrm{MV}}^l$ & $-0.12\%$ & $-0.13\%$ & $-0.15\%$ & $-0.10\%$ & $-0.76\%$ & $0.25\%$ \\
$(S_{\mathrm{MV}}^l)_\omega$ & $\mathbf{-0.01\%}$ & $\mathbf{-0.03\%}$ & $\mathbf{-0.06\%}$ & $\mathbf{-0.02\%}$ & $-0.69\%$ & $\mathbf{0.16\%}$ \\
$(S_{\mathrm{MV}}^l)_{\omega^*}$ & $-0.03\%$ & $+0.06\%$ & $+0.32\%$ & $+0.44\%$ & $\mathbf{-0.29\%}$ & $0.23\%$ \\
MC & $20.4882$ & $23.1697$ & $26.8056$ & $29.6348$ & $36.8715$ & NA \\
\hline
\end{tabular}
\caption{Approximations for selected quantiles of $S$ based on $S_{\mathrm{MV}}^l$ with $\mu=0.075$, $\sigma=0.15$, $n=20$, and yearly payments of 1.}
\label{tab:MV_quantiles}
\end{table}

\begin{table}[htbp]
\centering
\begin{tabular}{lrrrrrr}
\hline
Method & $p=0.25$ & $p=0.5$ & $p=0.75$ & $p=0.9$ & $p=0.95$ & MAD \\
\hline
$S_{\mathrm{MV}}^l$ & $-0.05\%$ & $-0.07\%$ & $-0.09\%$ & $-0.14\%$ & $-0.19\%$ & $0.11\%$ \\
$(S_{\mathrm{MV}}^l)_\omega$ & $\mathbf{-0.02\%}$ & $+0.05\%$ & $+0.03\%$ & $-0.04\%$ & $-0.10\%$ & $0.05\%$ \\
$(S_{\mathrm{MV}}^l)_{\omega^*}$ & $-0.04\%$ & $\mathbf{-0.03\%}$ & $\mathbf{\textcolor{green}{-0.01\%}}$ & $\mathbf{+0.04\%}$ & $\mathbf{+0.08\%}$ & $\mathbf{0.04\%}$ \\
MC & $13.5965$ & $15.3713$ & $18.1574$ & $21.7366$ & $24.4722$ & NA \\
\hline
Method & $p=0.95$ & $p=0.975$ & $p=0.99$ & $p=0.995$ & $p=0.999$ & MAD \\
\hline
$S_{\mathrm{MV}}^l$ & $-0.19\%$ & $-0.28\%$ & $-0.41\%$ & $-0.66\%$ & $-1.66\%$ & $0.64\%$ \\
$(S_{\mathrm{MV}}^l)_\omega$ & $-0.10\%$ & $-0.19\%$ & $-0.33\%$ & $-0.58\%$ & $-1.59\%$ & $0.56\%$ \\
$(S_{\mathrm{MV}}^l)_{\omega^*}$ & $\mathbf{+0.08\%}$ & $\mathbf{+0.11\%}$ & $\mathbf{+0.11\%}$ & $\mathbf{-0.16\%}$ & $\mathbf{-1.21\%}$ & $\mathbf{0.34\%}$ \\
MC & $24.4722$ & $27.2812$ & $31.1339$ & $34.2242$ & $42.0901$ & NA \\
\hline
\end{tabular}
\caption{Approximations for selected CTE’s of $S$ based on $S_{\mathrm{MV}}^l$ with $\mu=0.075$, $\sigma=0.15$, $n=20$, and yearly payments of 1.}
\label{tab:MV_CTE}
\end{table}

The MV example further demonstrates the advantages of combining exact moment matching with the minimum-deformation principle. Both the equally spaced and minimum-deformation corrections eliminate the variance discrepancy of the original MV approximation by matching the first two moments of $S$. However, the minimum-deformation correction achieves this with a substantially smaller Pearson deformation and, in the present example, also yields more accurate CTE approximations. Although the equally spaced correction performs better for the upper quantiles, the minimum-deformation correction retains its advantage in terms of Pearson deformation. These findings highlight the distinction between minimizing distributional deformation and minimizing approximation errors, while demonstrating that the proposed method can achieve exact moment matching without sacrificing the high accuracy of the starting MV approximation.

\subsection{Higher-order moment matching} \label{116870}
In the preceding subsections, we considered minimum-deformation comonotonic approximations matching the first two moments of $S$. In this subsection, we extend the numerical investigation to higher-order moment matching and examine the influence of the minimum interval-width parameter $\delta$ on the Pearson deformation. For illustration, we use the MV lower bound $S^l_{MV}$ as the starting comonotonic approximation and retain the model parameters specified in Subsection \ref{1030758}.

We consider $r=2,3,4,5$ and set $m=r+1$ for each moment-matching order. In this case, the correction weights are uniquely determined by the partition points through the moment-matching equations and the joint optimization problem can be reduced to an optimization problem over the partition points, subject to the nonnegativity of the weights and the minimum interval-width constraints. For each moment-matching order, we consider four values of $\delta$, namely $0.001$, $0.0001$, $0.00001$, and $0.000001$. In the numerical experiments below, feasible partitions are obtained for all the cases considered, so the corresponding minimum-deformation approximations exist. 

Table~\ref{tab:higher_order_deformation} reports the corresponding Pearson deformations. The results show that the choice of $\delta$ can have a substantial influence on the Pearson deformation, particularly for higher-order moment matching. For $r=2$, the Pearson deformation remains unchanged at the reported precision as $\delta$ decreases from $10^{-3}$ to $10^{-6}$. For $r=3$, only a slight reduction is observed. In contrast, for $r=4$, the deformation decreases by approximately $72.08\%$ as $\delta$ decreases from $10^{-3}$ to $10^{-6}$. For $r=5$, reducing $\delta$ from $10^{-3}$ to $10^{-4}$ decreases the deformation by approximately $\textcolor{green}{76.74\%}$, with no further reduction observed in the subsequent numerical results.

\begin{table}[htbp]
\centering
\begin{tabular}{ccccc}
\hline
$r$ & $\delta=10^{-3}$ & $\delta=10^{-4}$ & $\delta=10^{-5}$ & $\delta=10^{-6}$ \\
\hline
2 & 0.00001814 & 0.00001814 & 0.00001814 & 0.00001814 \\
3 & 0.00001508 & 0.00001497 & 0.00001497 & 0.00001497 \\
4 & 0.00006522 & 0.00002319 & 0.00001880 & 0.00001821 \\
5 & \textcolor{green}{0.00011342} & 0.00002639 & 0.00002639 & 0.00002639 \\
\hline
\end{tabular}
\caption{Pearson deformations for different values of $r$ and $\delta$.}
\label{tab:higher_order_deformation}
\end{table}

Table~\ref{tab:higher_order_constraints} indicates whether the minimum interval-width constraint is active at the numerically obtained solutions. For $r=2$, the constraint is inactive for all four values of $\delta$. For $r=3$ and $r=5$, it is active only at $\delta=10^{-3}$. For $r=4$, the constraint is active at $\delta=10^{-3},10^{-4},10^{-5}$, but inactive at $\delta=10^{-6}$. These observations indicate that the minimum interval-width constraint can play an important role in the construction of higher-order moment-matching approximations. In particular, when the constraint is active, reducing $\delta$ enlarges the feasible set and may lead to a smaller Pearson deformation. This behavior is particularly noticeable for $r=4$, where the deformation decreases as $\delta$ is reduced.

\begin{table}[htbp]
\centering
\begin{tabular}{ccccc}
\hline
$r$ & $\delta=10^{-3}$ & $\delta=10^{-4}$ & $\delta=10^{-5}$ & $\delta=10^{-6}$ \\
\hline
2 & No & No & No & No \\
3 & Yes & No & No & No \\
4 & Yes & Yes & Yes & No \\
5 & Yes & No & No & No \\
\hline
\end{tabular}
\caption{Activation of the minimum interval-width constraints.}
\label{tab:higher_order_constraints}
\end{table}

Overall, these numerical results illustrate the construction of step-weighted comonotonic approximations matching up to the first five moments of $S$ and demonstrate the influence of the minimum interval-width parameter on the Pearson deformation. The results also suggest that the choice of $\delta$ deserves particular attention in higher-order moment matching, especially when the numerically obtained partition contains very narrow intervals in the upper tail.

\section{Concluding remarks} \label{1031960}
In this paper, we have developed a step-weighted approach to constructing moment-matching comonotonic approximations for sums $S$ of lognormal random variables. By appropriately selecting the partition points and correction weights, the resulting approximations preserve the comonotonic structure while matching prescribed moments of $S$. The compatibility of step-weighted corrections with comonotonic summation provides a theoretical basis for this construction.

To determine the partition points and correction weights, we have introduced a minimum-deformation criterion based on the Pearson deformation. We have considered both fixed-partition and joint optimization problems and established results concerning the existence and uniqueness of their optimal solutions. The fixed-partition problem determines the optimal correction weights for prescribed partition points, whereas the joint optimization problem optimizes the partition points and correction weights simultaneously.

In particular, when the number of steps satisfies \(m=r+1\), we have shown that the correction weights are uniquely determined by the partition points through the moment-matching equations. Consequently, the joint optimization problem can be reduced to an optimization problem involving only the partition points, subject to the nonnegativity of the correction weights and the minimum interval-width constraints. This reduction eliminates the correction weights from the optimization variables and simplifies the formulation of the joint problem.

We have then applied the general results to sums of lognormal random variables. In this setting, we have established the existence of minimum-deformation comonotonic approximations whenever the corresponding feasible partition set is nonempty. In particular, for \(r=2\) and \(m=3\), we have derived sufficient conditions ensuring the nonemptiness of the feasible partition set and hence the existence of minimum-deformation comonotonic approximations for sufficiently small values of the minimum interval-width parameter \(\delta\). These results provide a theoretical foundation for constructing minimum-deformation comonotonic approximations for sums of lognormal random variables.

Finally, numerical experiments have illustrated the proposed approach using different starting comonotonic approximations. The results demonstrate the influence of the starting approximation on the approximation performance of the resulting moment-matching approximation. In particular, minimizing the Pearson deformation does not necessarily lead to improved approximation accuracy. Further experiments involving higher-order moment matching have shown that the value of \(\delta\) can have a substantial influence on the Pearson deformation. These findings suggest that careful selection of both the starting comonotonic approximation and the partition points is important in constructing minimum-deformation comonotonic approximations. Further investigation of the efficiency and stability of the proposed optimization approach, particularly for higher-order moment matching, would be of interest.


\begin{thebibliography}{99}
\bibitem{AS1966}
Ali S M, Silvey S D. A general class of coefficients of divergence of one distribution from another[J]. Journal of the Royal Statistical Society: Series B (Methodological), 1966, 28(1): 131-142.

\bibitem{BKA1998}
Baggerly K A. Empirical likelihood as a goodness-of-fit measure[J]. Biometrika, 1998, 85(3): 535-547.
\bibitem{BMKA2012}
Broniatowski M, Keziou A. Divergences and duality for estimation and test under moment condition models[J]. Journal of Statistical Planning and Inference, 2012, 142(9): 2554-2573.





\bibitem{DDV2011}
Deelstra G, Dhaene J, Vanmaele M. An overview of comonotonicity and its applications in finance and insurance[J]. Advanced mathematical methods for finance, 2011: 155-179. 

\bibitem{DD2007}
Denuit M, Dhaene J. Comonotonic bounds on the survival probabilities in the Lee–Carter model for mortality projection[J]. Journal of Computational and Applied Mathematics, 2007, 203(1): 169-176.

\bibitem{DDGK2006}
Denuit M, Dhaene J, Goovaerts M, et al. Actuarial theory for dependent risks: measures, orders and models[M]. John Wiley \& Sons, 2006.

\bibitem{Denuit2019}
Denuit M. Size-biased transform and conditional mean risk sharing, with application to P2P insurance and tontines[J]. ASTIN Bulletin: The Journal of the IAA, 2019, 49(3): 591-617.

\bibitem{Denuit2020}
Denuit M. Size-biased risk measures of compound sums[J]. North American Actuarial Journal, 2020, 24(4): 512-532.

\bibitem{Dhaene et al. (2002a)}
Dhaene J, Denuit M, Goovaerts M J, et al. The concept of comonotonicity in actuarial science and finance: theory[J]. Insurance: Mathematics and Economics, 2002, 31(1): 3-33. 

\bibitem{Dhaene et al. (2002b)}
Dhaene J, Denuit M, Goovaerts M J, et al. The concept of comonotonicity in actuarial science and finance: applications[J]. Insurance: Mathematics and Economics, 2002, 31(2): 133-161. 

\bibitem{DVG2005}
Dhaene J, Vanduffel S, Goovaerts M J, et al. Comonotonic approximations for optimal portfolio selection problems[J]. Journal of Risk and Insurance, 2005, 72(2): 253-300.

\bibitem{Dhaene et al. (2006)}
Dhaene J, Vanduffel S, Goovaerts M J, et al. Risk measures and comonotonicity: a review[J]. Stochastic models, 2006, 22(4): 573-606.

\bibitem{DHL2008}
Dhaene J, Henrard L, Landsman Z, et al. Some results on the CTE-based capital allocation rule[J]. Insurance: Mathematics and Economics, 2008, 42(2): 855-863.

\bibitem{DWY2000}
Dhaene J, Wang S, Young V R, et al. Comonotonicity and maximal stop-loss premiums[J]. Bulletin of the Swiss Association of Actuaries, 2000, 2: 99-113.

\bibitem{FZ2008a}
Furman E, Zitikis R. Weighted premium calculation principles[J]. Insurance: Mathematics and Economics, 2008, 42(1): 459-465.

\bibitem{FZ2008b}
Furman E, Zitikis R. Weighted risk capital allocations[J]. Insurance: Mathematics and Economics, 2008, 43(2): 263-269.

\bibitem{FZ2009}
Furman E, Zitikis R. Weighted pricing functionals with applications to insurance: an overview[J]. North American Actuarial Journal, 2009, 13(4): 483-496.

\bibitem{FL2005}
Furman E, Landsman Z. Risk capital decomposition for a multivariate dependent gamma portfolio[J]. Insurance: Mathematics and Economics, 2005, 37(3): 635-649.

\bibitem{FL2008}
Furman E, Landsman Z. Economic capital allocations for non-negative portfolios of dependent risks[J]. ASTIN Bulletin: The Journal of the IAA, 2008, 38(2): 601-619.

\bibitem{HuangStep}
C. Huang, Comonotonic and moment matching approximations for sums of lognormal random variables, arXiv preprint arXiv:2606.29143, 2026.

\bibitem{HuangDRM}
C. Huang, Distortion risk measures of step-weighted distribution, arXiv preprint arXiv:2607.09132, 2026.

\bibitem{KDG2000}
Kaas R, Dhaene J, Goovaerts M J. Upper and lower bounds for sums of random variables[J]. Insurance: Mathematics and Economics, 2000, 27(2): 151-168.

\bibitem{KGDD2008}
Kaas R, Goovaerts M, Dhaene J, et al. Modern actuarial risk theory: using R[M]. Berlin, Heidelberg: Springer Berlin Heidelberg, 2008.

\bibitem{KPW2012}
Klugman S A, Panjer H H, Willmot G E. Loss models: from data to decisions[M]. John Wiley \& Sons, 2012.

\bibitem{LV2006}
Liese F, Vajda I. On divergences and informations in statistics and information theory[J]. IEEE Transactions on Information Theory, 2006, 52(10): 4394-4412.

\bibitem{MP1998}
Milevsky M A, Posner S E. Asian options, the sum of lognormals, and the reciprocal gamma distribution[J]. Journal of financial and quantitative analysis, 1998, 33(3): 409-422.

\bibitem{MR2000}
Milevsky M A, Robinson C. Self-annuitization and ruin in retirement[J]. North American Actuarial Journal, 2000, 4(4): 112-124.

\bibitem{PR1976}
Patil G P, Ord J K. On size-biased sampling and related form-invariant weighted distributions[J]. Sankhyā: The Indian Journal of Statistics, Series B, 1976: 48-61.

\bibitem{PR1978}
Patil G P, Rao C R. Weighted distributions and size-biased sampling with applications to wildlife populations and human families[J]. Biometrics, 1978: 179-189.

\bibitem{VCD2008}
Vanduffel S, Chen X, Dhaene J, et al. Optimal approximations for risk measures of sums of lognormals based on conditional expectations[J]. Journal of Computational and Applied Mathematics, 2008, 221(1): 202-218.

\bibitem{VDG2005a}
Vanduffel S, Dhaene J, Goovaerts M. On the evaluation of ‘saving-consumption’plans[J]. Journal of Pension Economics \& Finance, 2005, 4(1): 17-30. 

\bibitem{VDG2003}
Vanduffel S, Dhaene J, Goovaerts M, et al. The hurdle-race problem[J]. Insurance: Mathematics and Economics, 2003, 33(2): 405-413.

\bibitem{VHD2005b} 
Vanduffel S, Hoedemakers T, Dhaene J. Comparing approximations for risk measures of sums of nonindependent lognormal random variables[J]. North American Actuarial Journal, 2005, 9(4): 71-82. 

\bibitem{WD1998}
Wang S, Dhaene J. Comonotonicity, correlation order and premium principles[J]. Insurance: Mathematics and Economics, 1998, 22(3): 235-242. 
\end{thebibliography}
\end{document}